\documentclass[apj]{emulateapj}
\usepackage{graphicx}

\def\HI{{\sc Hi}}
\def\HII{{\sc Hii}}
\def\Ep{E_\mathrm{p}}
\def\Msol{M_\odot}
\shorttitle{Kinematic distance study of the complex G35.6-0.4}
\shortauthors{Zhu et al.}

\begin{document}

\title{Kinematic distance study of the planetary nebulae-supernova \\ remnant-{\HII} region complex at G35.6-0.5}

\author{H. Zhu\altaffilmark{1}, W. W. Tian\altaffilmark{1,2}, D. F. Torres\altaffilmark{3,4}, G. Pedaletti\altaffilmark{3} \& H. Q. Su\altaffilmark{1}}
\affil{$^1$Key Laboratory of Optical Astronomy, National Astronomical Observatories, Chinese Academy of Sciences,
Beijing 100012, China; zhuhui@bao.ac.cn and tww@bao.ac.cn}
\affil{$^2$Department of Physics \& Astronomy, University of Calgary, Calgary, Alberta T2N 1N4, Canada}
\affil{$^3$Institute of Space Sciences (IEEC-CSIC), Campus UAB,  Torre C5, 2a planta, 08193 Barcelona, Spain}
\affil{$^4$Instituci\'o Catalana de Recerca i Estudis Avan\c{c}ats (ICREA) Barcelona, Spain}

\begin{abstract}
Two possible planetary nebulae (PN G035.5-00.4 and IRAS 18551+0159), one newly re-identified supernova remnant (SNR G35.6-0.4), and one {\HII} region (G35.6-0.5) form a line-of-sight-overlapped complex known as G35.6-0.5. We analyze 21 cm {\HI} absorption spectra towards the complex to constrain their kinematic distances. PN G035.5-00.4 has a distance from 3.8$\pm$0.4 kpc to 5.4$\pm$0.7 kpc. IRAS 18551+0159 is at 4.3$\pm$0.5 kpc. We discuss the distance for SNR 35.6-0.4, for which the previous estimate was 10.5 kpc, and find plausible for it to be 3.6$\pm$0.4 kpc. The new distance of SNR G35.6-0.4 and the derived mass for the $\sim55$ km s${}^{-1}$ CO molecular cloud can accommodate an association with HESS J1858+020. We also conclude that SNR G35.6-0.4 is unlikely associated with PSR J1857+0210 or PSR J1857+0212, which are projected into the SNR area.
\end{abstract}

\keywords{{\HII} regions --- planetary nebulae: individual (PN G035.5-00.4, IRAS 18551+0159) --- supernova remnants}

\section{Introduction}
\noindent supernova remnants (SNRs) and {\HII} regions are two classes of bright radio objects in Galactic plane. They play a key role in understanding the structure and evolution of the Milky Way. Planetary nebulae (PNe) are important probe of nucleosynthesis processes and responsible for a large fraction of the chemical enrichment of interstellar medium. The complex G35.6-0.5, near Galactic coordinates l=$35.6^{\circ}$ b=$-0.5^{\circ}$, consists of SNR G35.6-0.4, PNe (PN G035.5-00.4, IRAS 18551+0159) and {\HII} region G35.6-0.5 (see Figure \ref{fig:complex}). This region has been studied since 1969 (\citealp*[e.g.][]{bea69,loc89,kuc90}; \citealt{par06,kwo08}; \citealp*{par10}). However, some basic issues related to the complex are still unclear.

PN G035.5-00.4 and IRAS 18551+0159 were discovered and identified as possible PNe by \citet*{kis95}. Their angular sizes are about $10{''}$ and $3{''}$ respectively, based on infrared and radio observations (\citealt{kwo08}; \citealp*{phi09}). IRAS 18551+0159 is extremely red and has no optical counterpart. This implies IRAS 18551+0159 just left from the Asymptotic Giant Branch (AGB) recently (\citealt{jim05}). \citet*{loc89} and \citet*{phi93} detected radio recombination lines (RRLs) at a position of l=$35.588^{\circ}$ b=$-0.489^{\circ}$, close to PN G035.5-00.4. They identified this RRLs source as a likely {\HII} region G35.6-0.5. \citet*{gre09} questioned the existence of the {\HII} region G35.6-0.5 because there is no 100 $\mu$m emission from it and the RRLs may come from the nearby PN G035.5-00.4.

SNR G35.6-0.4 has the most tortuous discovery history. It had first been identified as an SNR due to its deep spectral index derived by both \citet*{vel74} and \citet*{dic75}. \citet{ang77} also suggested that G35.6-0.4 is an SNR. However, there existed opposing claims, e.g. this source has a flat spectral index \citep*{cas75}; RRLs is detected within the source. However, \citet{gre09} reanalyzed previous measurements and derived a deep spectral index to G35.6-0.4. Moreover, the IRAS 100 $\mu$m image showed in his Figure 3 is a supporting evidence that G35.6-0.4 is an SNR.

\citet*{phi93} suggested a kinematic distance of 12 kpc for SNR G35.6-0.4, which consists with the dispersion measure (DM) distance of $\sim$13 kpc to PSR J1857+0212. Therefore, they thought that the SNR G35.6-0.4/PSR J1857+0212 association was real. Although the recent-estimated DM distance to PSR J1857+0212 has been changed to 7.98 kpc \citep{han06}, but the SNR changed its distance too, i.e. 10.5 kpc \citep[assuming that SNR G35.6-0.4 is at the same distance as the distance-known {\HII} region G35.5-0.0,][]{gre09}. Thus, the SNR G35.6-0.4/PSR J1857+0212 association seems still possible according to recent studies.

\citet*{par10} noticed that the weak gamma ray source HESS J1858+020 is located at the southern border of SNR G35.6-0.4 (see Figure 1 of their paper). They analyzed the ${}^{13}\textrm{CO J=1-0}$ line of two molecular clouds (MCs) towards the southern border of G35.6-0.4, and suggested the association between the SNR G35.6-0.4 and the $\sim$55 km s${}^{-1}$ MC as the counterpart of HESS J1858+020. \citet{tor11} pointed out if this association is real, i.e., the HESS J1858+020 is connected with cosmic-ray protons accelerated by SNR G35.6-0.4, it would be difficult not to produce detectable GeV emission at the same location. These results are based on a distance of 10.5 kpc to both SNR G35.6-0.4 and the CO cloud and the derived molecular mass in the environment.

It is important to obtain reliable distances of the components of complex G35.6-0.5 to clarify their relationships (\citealp*[e.g. the W51 complex,][]{tia13}; \citealt{bro13}). In this paper, we measure the kinematic distance of the complex using 1420 MHz continuum, 21 cm {\HI} spectral line and ${}^{13}\textrm{CO J=1-0}$ line data. We also discuss the relationships among its components.

\section{Data and method to build absorption spectrum}

\subsection{Data}

\noindent The 1420 MHz radio continuum and {\HI} emission data come from the Very Large Array (VLA) Galactic Plane Survey \citep[VGPS,][]{sti06}. The continuum images have a spatial resolution of $1{'}$ at 1420 MHz. The {\HI} spectral line images, shown with a resolution of $1{'}\times$$1{'}\times$ 1.56 km s${}^{-1}$, have an rms noise of 2 K per 0.824 km s${}^{-1}$ channel. The ${}^{13}\textrm{CO} (J=1-0)$ spectral line data is from the Galactic Ring Survey implemented with the Five College Radio Observatory 14 m telescope \citep{jac06}. These data have an angular and spectral resolution of $46{''}$ and 0.21 km s${}^{-1}$ respectively.

\subsection{The method to build reliable absorption spectrum}

\noindent We use the following formulas to calculate the absorption spectrum:
\newline
For the source (ON),\\
\begin{equation}
{T_{on}}(v) = {T_B}(v)(1 - {e^{ - {\tau _t}(v)}}) + T_s^c({e^{ - {\tau _c}(v)}} - 1),
\end{equation}
\newline
For the background (OFF),\\
\begin{equation}
{T_{off}}(v) = {T_B}(v)(1 - {e^{ - {\tau _t}(v)}}) + T_{bg}^c({e^{ - {\tau _c}(v)}} - 1).
\end{equation}
\newline
Then we get the expression of 21 cm absorption spectrum,\\
\begin{equation}
{{e^{ - {\tau _c}(v)}} = 1 - \frac{{{T_{off}}(v) - {T_{on}}(v)}}{{T_s^c - T_{bg}^c}}}.
\end{equation}
\newline
The `-1' in the second term of equation (1) \& (2) refers to the subtracted continuum emission. ${T_{on}}(v)$ and ${T_{off}}(v)$ are the {\HI} brightness temperatures of the source and the background regions at the velocity of $v$. $T_s^c$ and $T_{bg}^c$ are the continuum brightness temperatures for the regions same as ${T_{on}}(v)$ and ${T_{off}}(v)$. ${T_B}(v)$ is the spin temperature of the {\HI} cloud.

To build the {\HI} absorption spectrum, traditional method selects the background region ($T_{bg}^c$) separated from the continuum source ($T_s^c$). However, there is a possibility to construct a false absorption spectrum due to the potential difference in the {\HI} distribution along two lines of sight. \citet{tia07} presented a revised method to build 21 cm {\HI} absorption spectrum against a background extended source. In their method, the background region surrounds the source region directly. This minimizes the possibility of obtaining a false absorption spectrum. In addition, they use CO spectrum in the source direction, and the {\HI} absorption spectrum of other bright continuum sources nearby, to understand the absorption spectrum of the target source. In this paper, the minimum standard for a reliable absorption feature is taken as
${{e^{ - {\tau _c}(v)}}} > {1 - \frac{{3\Delta T}}{{T_s^c - T_{bg}^c}}}$ (used as $3\sigma$), where ${\Delta T}$ is calculated based on baselines without emissions and used as an estimate of fluctuations caused by the receiver noise.

\section{Results and analysis}

\subsection{{\HI} absorption and ${}^{13}\textrm{CO}$ emission spectra}

\begin{figure*}[!ht]
\centerline{\includegraphics[width=0.5\textwidth, angle=0]{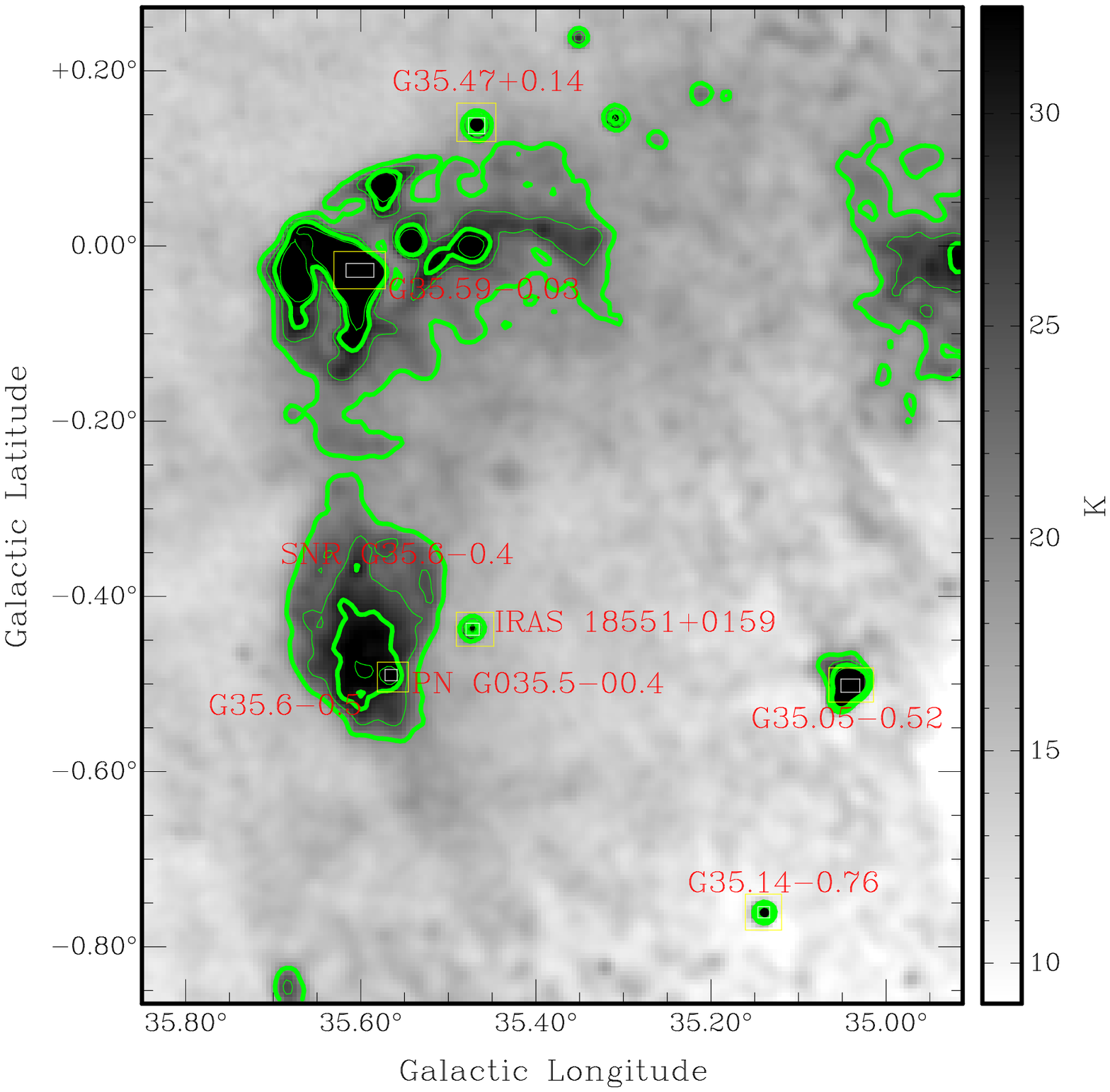}\includegraphics[width=0.5\textwidth, angle=0]{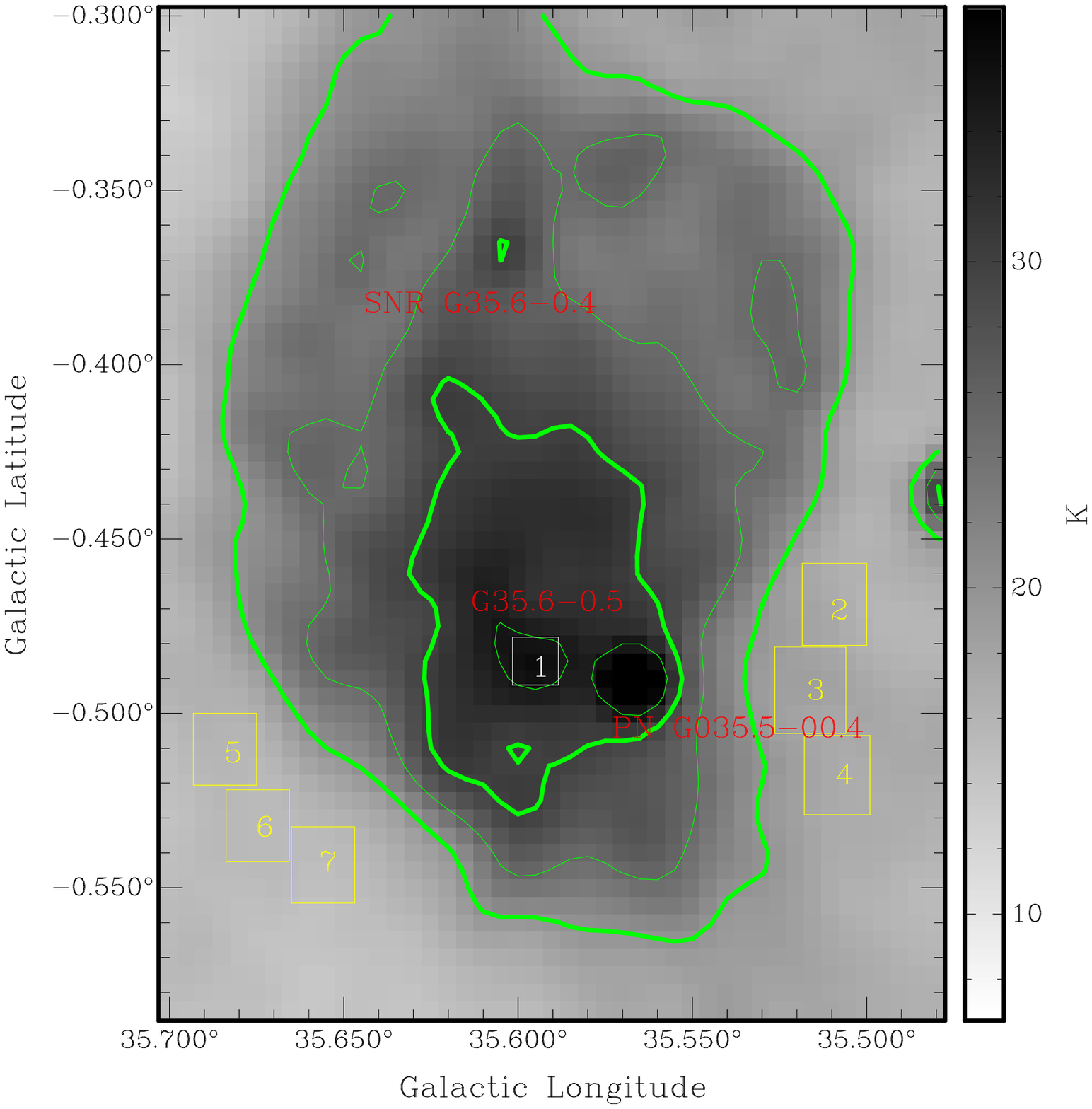}}\caption{Left: 1420 MHz continuum image of the complex, including PNe: PN G035.5-00.4, IRAS 18551+0159; SNR: SNR G35.6-0.4; {\HII} regions: G35.6-0.5, G35.47+0.14, G35.59-0.03, G35.05-0.52, G35.14-0.76. Right: A close-up on the PNe-SNR-{\HII} region complex. These green contours delineate the radio emission with levels of 20K, 25K, 30K \& 35K. White boxes are used as source regions, while yellow boxes are used as background regions.}
\label{fig:complex}
\end{figure*}

\noindent The left panel of Figure \ref{fig:complex} shows the 1420 MHz continuum image of the complex with contours (20, 25, 30 and 35 K), including PNe: PN G035.5-00.4 and IRAS 18551+0159, SNR G35.6-0.4, {\HII} regions: G35.47+0.14 \citep[an ultracompact {\HII} region,][]{giv07}, G35.6-0.5, G35.59-0.03, G35.05-0.52 and G35.14-0.76 (re-identified as {\HII} region by Froebrich \& Ioannidis 2011). The right panel of Figure \ref{fig:complex} displays a close-up of the complex with an angular size of $\sim13{'}$$\times 17{'}$. Two small bright areas are centered at \emph{l=$35.56\,^{\circ}$, b=$-0.49\,^{\circ}$} and \emph{l=$35.59\,^{\circ}$, b=$-0.49\,^{\circ}$}, the same coordinates as PN G035.5-00.4 and {\HII} region G35.6-0.5. For PN G035.5-00.4, IRAS 18551+0159 and the {\HII} regions except for G35.6-0.5, the method described by \citet{tia07} and \citet*{tia08} is used to extract the {\HI} spectrum. The source regions are presented by the white boxes in Figure \ref{fig:complex}. The background regions are the regions between the white box and the yellow box. Because SNR G35.6-0.4 and {\HII} region G35.6-0.5 are faint and extended, we use the traditional method to construct their {\HI} absorption spectra. The white box 1 is selected as their source region. The average of yellow boxes numbered from 2 to 7 is used as the background.

Figure \ref{fig:sourceabsorption} shows the {\HI} emission, absorption and CO emission spectra of PN G035.5-00.4 (the top panel), IRAS 18551+0159 (the middle panel) and SNR G35.6-0.4 and {\HII} region G35.6-0.5 (the bottom panel). Figure \ref{fig:backgroundabsorption} shows spectra of 4 {\HII} regions: G35.05-0.52 (top-left), G35.14-0.76 (top-right), G35.47+0.14 (bottom-left), and G35.59-0.03 (bottom-right). The lack of reliable {\HI} absorption at negative velocities for all sources in Figures \ref{fig:sourceabsorption} and \ref{fig:backgroundabsorption} imply that all of them are localized in the solar circle (For SNR G35.6-0.4 and {\HII} region G35.6-0.5, the $-10$ km s${}^{-1}$ and the $-46$ km s${}^{-1}$ absorption features are not real, see section 3.2.3). The nearly continuous absorption features from $\sim$10 km s${}^{-1}$ to $\sim$58 km s${}^{-1}$ are seen in all sources except for G35.14-0.76, indicating a lower limit to the distances of these objects. G35.14-0.76 only shows continuous absorption features up to $\sim$35 km s${}^{-1}$.

For PN G035.5-00.4, the highest {\HI} absorption velocity appears at $\sim$68 km s${}^{-1}$. However, the 68 km s${}^{-1}$ absorption feature disappears in the absorption spectrum averaged over 4 adjacent channels. Therefore, it is more reasonable to take $\sim$58 km s${}^{-1}$ as the highest absorption velocity which suggests a lower limit distance of PN G035.5-00.4. The ${}^{13}\textrm{CO}$ emission spectrum towards PN G035.5-00.4 shows a clear peak at $\sim$82 km s${}^{-1}$ without any associated {\HI} absorption. This hints that the CO cloud is behind PN G035.5-00.4. IRAS 18551+0159 shows its highest absorption feature at $\sim$66 km s${}^{-1}$. For SNR G35.6-0.4 and {\HII} region G35.6-0.5, the highest absorption velocity is at $\sim$61 km s${}^{-1}$. Like PN G035.5-00.4, there is no absorption at $\sim$82 km s${}^{-1}$, which is also confirmed by channel maps (see Figure \ref{fig:channelmap}). We detect absorption at the tangent point velocity of $\sim$106 km s${}^{-1}$ for G35.59-0.03 and G35.05-0.52. This implies that both of them beyond the tangent point and farther than the complex.

\subsection{Distances}
\noindent In this section, all kinematic distances are calculated based on a circular Galactic rotation curve model with the IAU adopted value ${{V_0} = 220}$ km s${}^{-1}$ and ${{R_0} = 8.5}$ kpc. The 7 km s${}^{-1}$ random motion \citep*{bel84} is employed to derive the distance uncertainty.
\subsubsection{Four {\HII} regions}
\noindent RRLs at 51.4$\pm$2.3 km s${}^{-1}$ and 51.2$\pm$1.9 km s${}^{-1}$ have been found towards {\HII} regions G35.59-0.03 and G35.05-0.52 respectively \citep*{loc89,kim01}. The velocities correspond to a near side distance of 3.4$\pm$0.4 kpc and a far side distance of 10.4$\pm$0.4 kpc. Because continuous {\HI} absorption features are seen from 51 km s${}^{-1}$ to the tangent point velocity, both {\HII} regions are at 10.4$\pm$0.4 kpc.

\citet{wat03} detected RRL at 80.9$\pm$0.5 km s${}^{-1}$ towards G35.47+0.14 which implies a near side distance of 5.3$\pm$0.7 kpc and a far side distance of 8.6$\pm$0.7 kpc. In the absorption spectrum towards G35.47+0.14, the highest absorption velocity is about 88 km s${}^{-1}$ (see the lower-left panel of Figure \ref{fig:backgroundabsorption}). There are no HI absorption features near the tangent point velocity although we find ${}^{13}\textrm{CO}$ emission peak at $\sim$101 km s${}^{-1}$. Therefore the near side distance of 5.3$\pm$0.7 kpc is a proper estimate to G35.47+0.14. The absorption features between 80.9 km s${}^{-1}$ and 88 km s${}^{-1}$ are likely caused by the random motion of HI clouds.

For G35.14-0.76, the highest {\HI} absorption velocity is $\sim$35 km s${}^{-1}$ at which there is a ${}^{13}\textrm{CO}$ emission peak. Compared with absorption spectra of three other {\HII} regions which show absorption features from 30 km s${}^{-1}$ to 50 km s${}^{-1}$, a distance of 2.4$\pm$0.5 kpc is rational to G35.14-0.76 (i.e. the near side distance of the 35 km s${}^{-1}$ absorption feature).

\begin{table*}
\begin{center}
\caption{Summary of HI absorption, CO emission features, and distances of the studied objects}
\setlength{\tabcolsep}{1mm}
\begin{tabular}{ccccccccc}
\hline
Name: & G35.59-0.03 & G35.05-0.52 & G35.47+0.14 & G35.14-0.76 & IRAS 18551+0159  & PN G035.5-00.5 & G35.6-0.5 & SNR G35.6-0.4 \\
\hline
\hline
MAV:   & $\sim$106 & $\sim$106 & $\sim$88 & $\sim$35 & $\sim$66 & $\sim$58 & $\sim$${61}^{a}$ & $\sim$${61}^{a}$ \\
RRLV: & 51.4$\pm$2.3 & 51.2$\pm$1.9 & 80.9$\pm$0.5 &  &  & ${55\pm3.6}^{b}$ & ${55\pm3.6}^{b}$ & \\
NCO:  &  &  & $\sim$82 & $\sim$35 &  & $\sim$58 & $\sim$${55}^{c}$ & $\sim$55\\
FCO:  &   &  & $\sim$101 & $\sim$44 & $\sim$82 & $\sim$82 & $\sim$82 & $\sim$82 \\
Distance: & 10.4$\pm$0.4 & 10.4$\pm$0.4 & 5.3$\pm$0.7 & 2.4$\pm$0.5 & 4.3$\pm$0.5 & 3.8$\pm$0.4 $\sim$ 5.4$\pm$0.7 & ${3.6\pm0.4}^{c}$ & 3.6$\pm$0.4 \\
\hline
\hline
\end{tabular}
\label{tbl:summary}
\end{center}
MAV: Maximum absorption velocity, RRLV: RRL velocity, NCO: velocity of nearby CO emission feature, FCO: velocity of far CO emission feature. The unit of velocity is km s${}^{-1}$. The unit of distance is kpc.\\
${}^{a}$: The SNR G35.6-0.4 overlaps with an {\HII} region G35.6-0.5, so we can not distinguish the highest absorption velocity (61 km s${}^{-1}$) feature from their mixed spectrum, e.g. one of the two sources' MAV might be less than 61 km s${}^{-1}$.\\
${}^{b}$: PN G035.5-00.4 and G35.6-0.5 are covered in the same RRL observation beam. It is unkonwn which one is responsible for the RRL.\\
${}^{c}$: This is based on the assumption that part of the RRL comes from G35.6-0.5.
\end{table*}

\subsubsection{Planetary nebulae: IRAS 18551+0159 and PN G035.5-00.4}

\noindent The highest absorption velocity of $\sim$66 km s${}^{-1}$ in the spectrum of IRAS 18551+0159 suggests a lower limit distance of 4.3$\pm$0.5 kpc to IRAS 18551+0159. In Figure 4., many HI self-absorption features at about 82 km s${}^{-1}$ are found around the complex consistent with the ${}^{13}\textrm{CO}$ emission at 82 km s${}^{-1}$ (red contour). The appearance of HI self-absorption features reveals that there are cold HI clouds at about 5.4 kpc (the near side distance of 82 km s${}^{-1}$). IRAS 18551+0159 is likely closer than 5.4$\pm$0.7 kpc because we do not find any absorption features at 82 km s${}^{-1}$ in its spectrum. Compared with the background source G35.47+0.14, which is not beyond the tangent point distance of 6.9 kpc and displays continuous absorption features from $\sim$70 km s${}^{-1}$ to $\sim$88 km s${}^{-1}$, we suggest a near distance of 4.3$\pm$0.5 kpc for IRAS 18551+0159.

\begin{figure}[!htpb]

   \includegraphics[width=0.5\textwidth, angle=0]{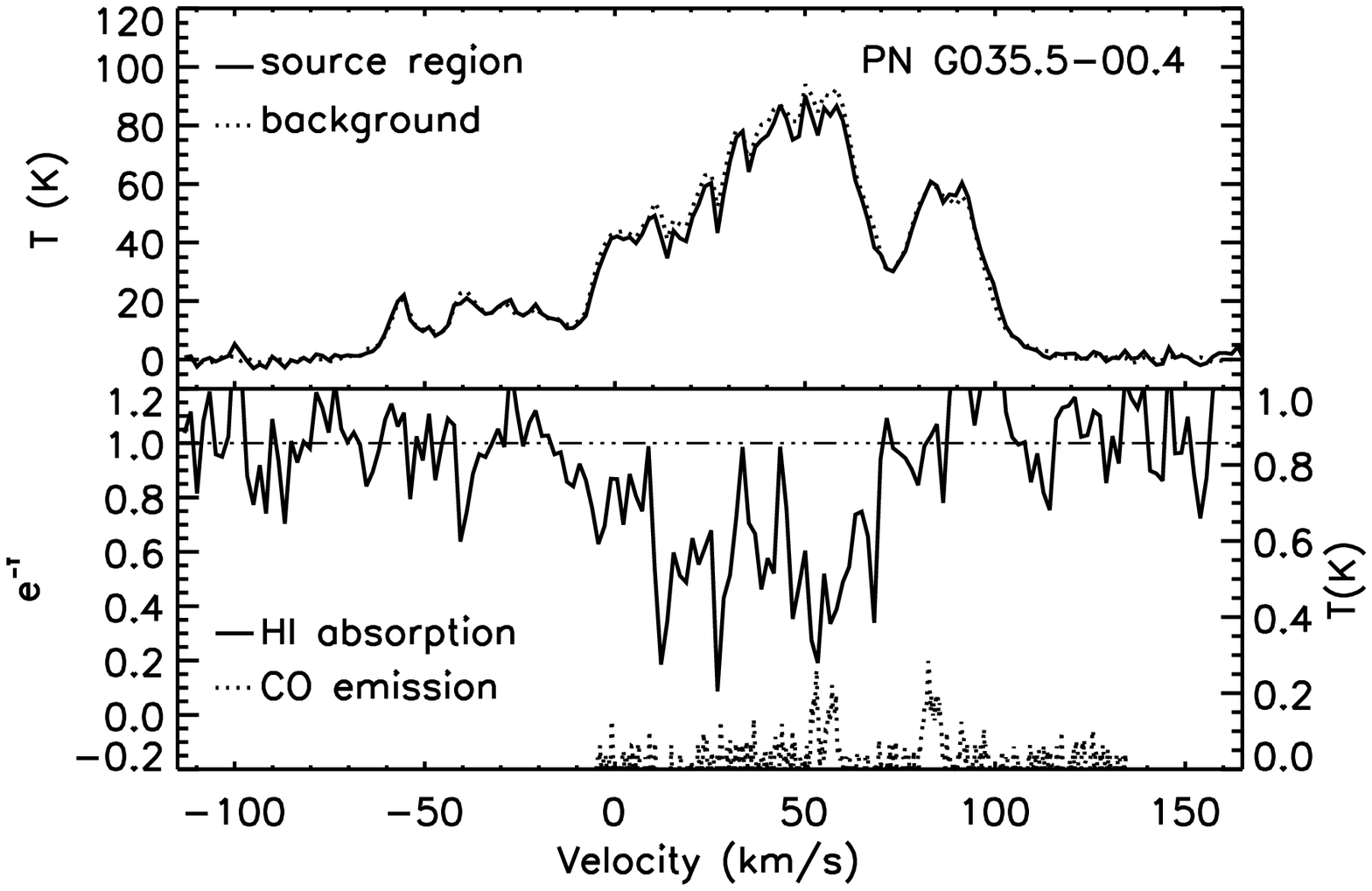}
   \includegraphics[width=0.5\textwidth, angle=0]{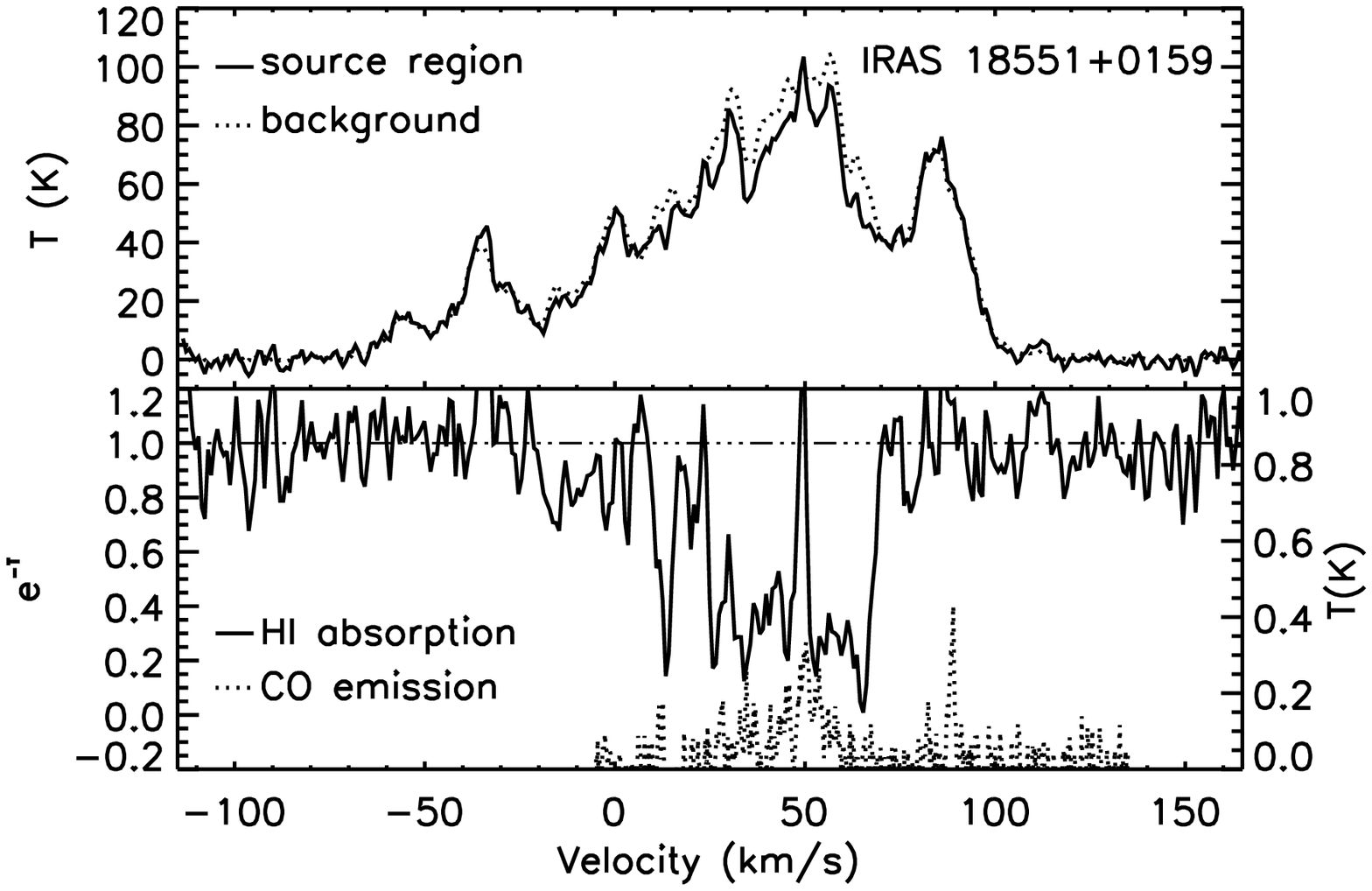}
   \includegraphics[width=0.5\textwidth, angle=0]{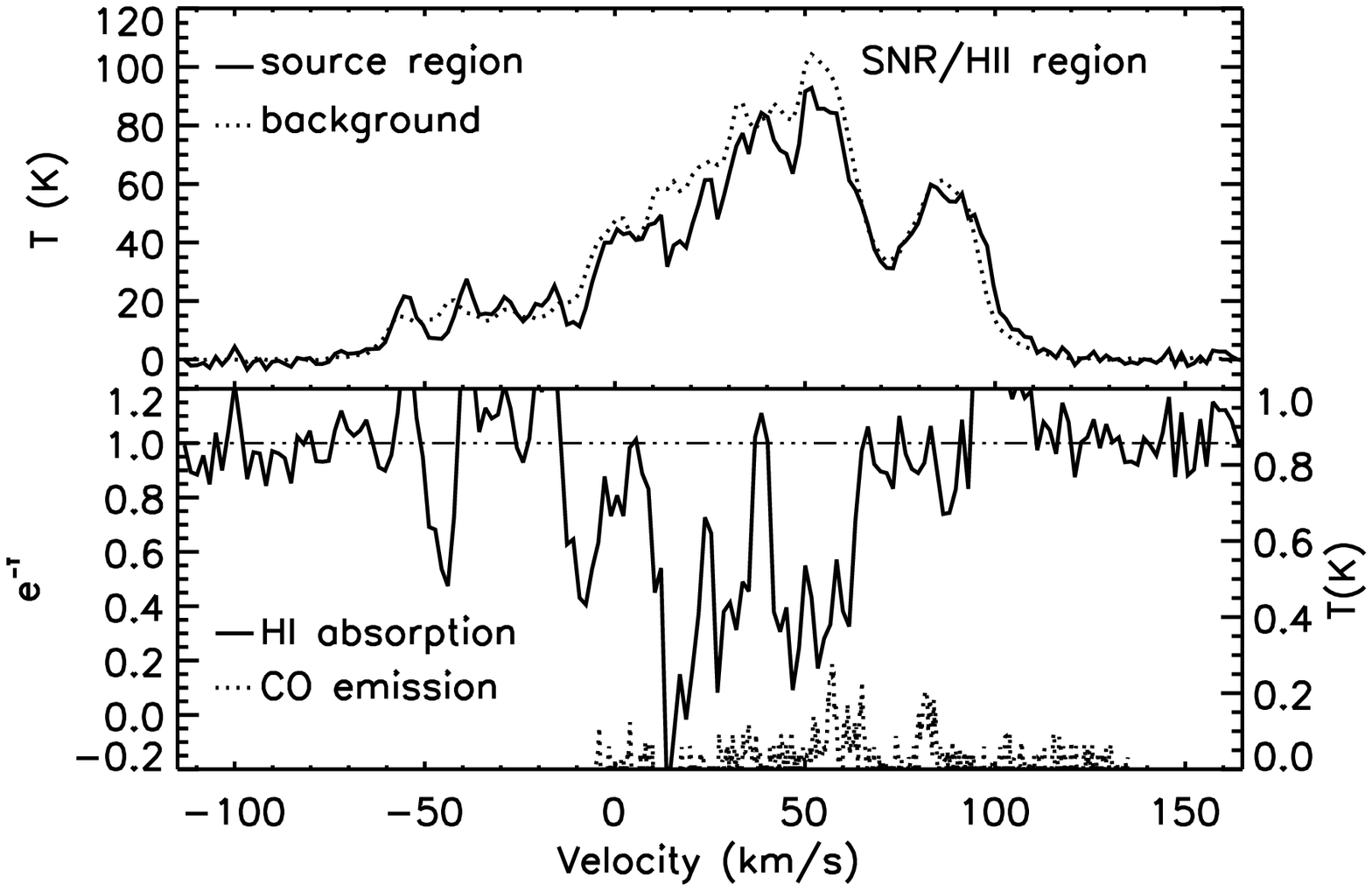}
   \caption{The {\HI}, ${}^{13}\textrm{CO}$ emission and {\HI} absorption spectra of PN G035.5-00.4 (the top panel), IRAS 18551+0159 (the middle panel) and SNR G35.6-0.4 and {\HII} region G35.6-0.5 (the bottom panel). }
   \label{fig:sourceabsorption}
   \end{figure}

\begin{figure*}[!htpb]
\centerline{\includegraphics[width=0.5\textwidth, angle=0]{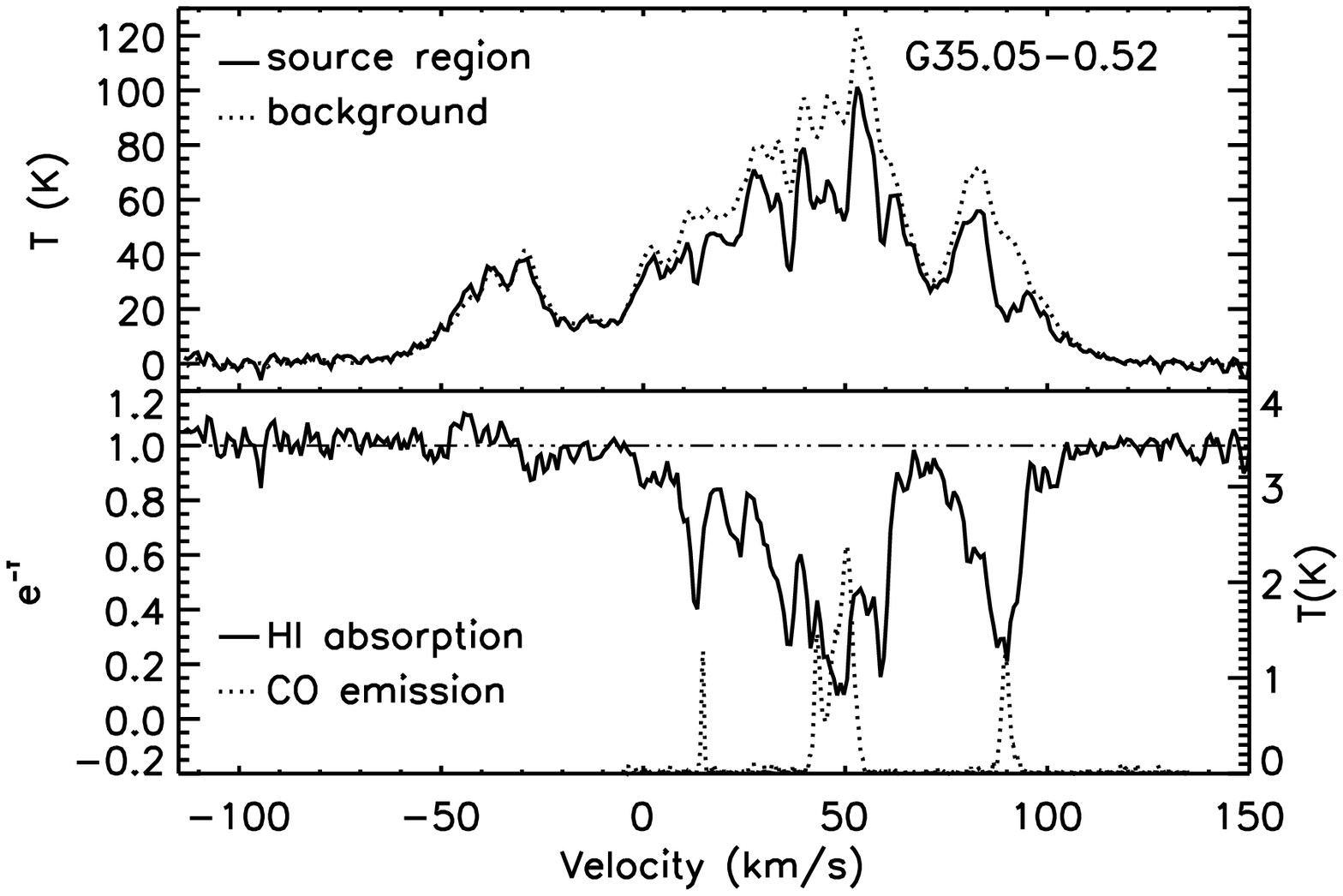}\includegraphics[width=0.5\textwidth, angle=0]{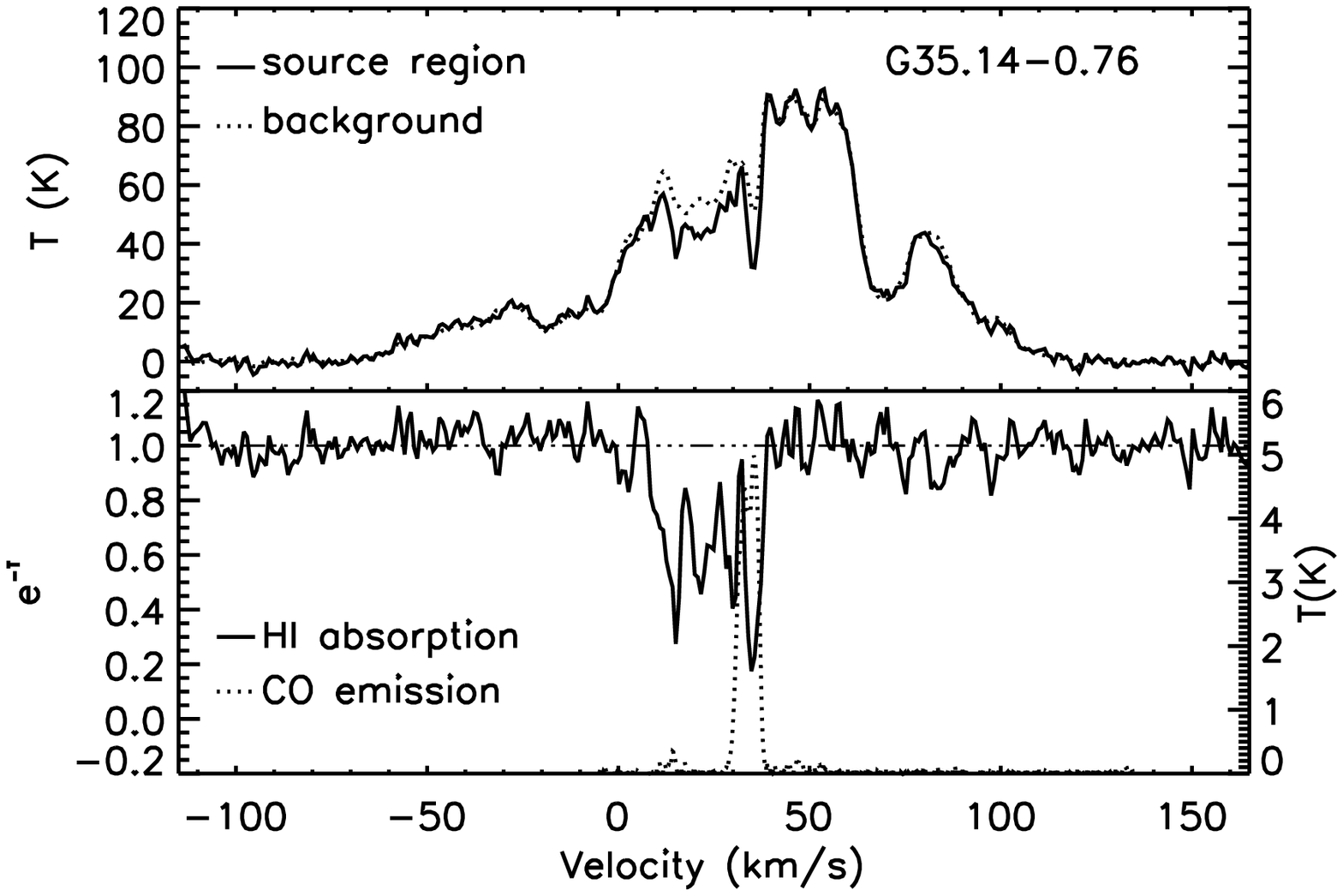}}
\centerline{\includegraphics[width=0.5\textwidth, angle=0]{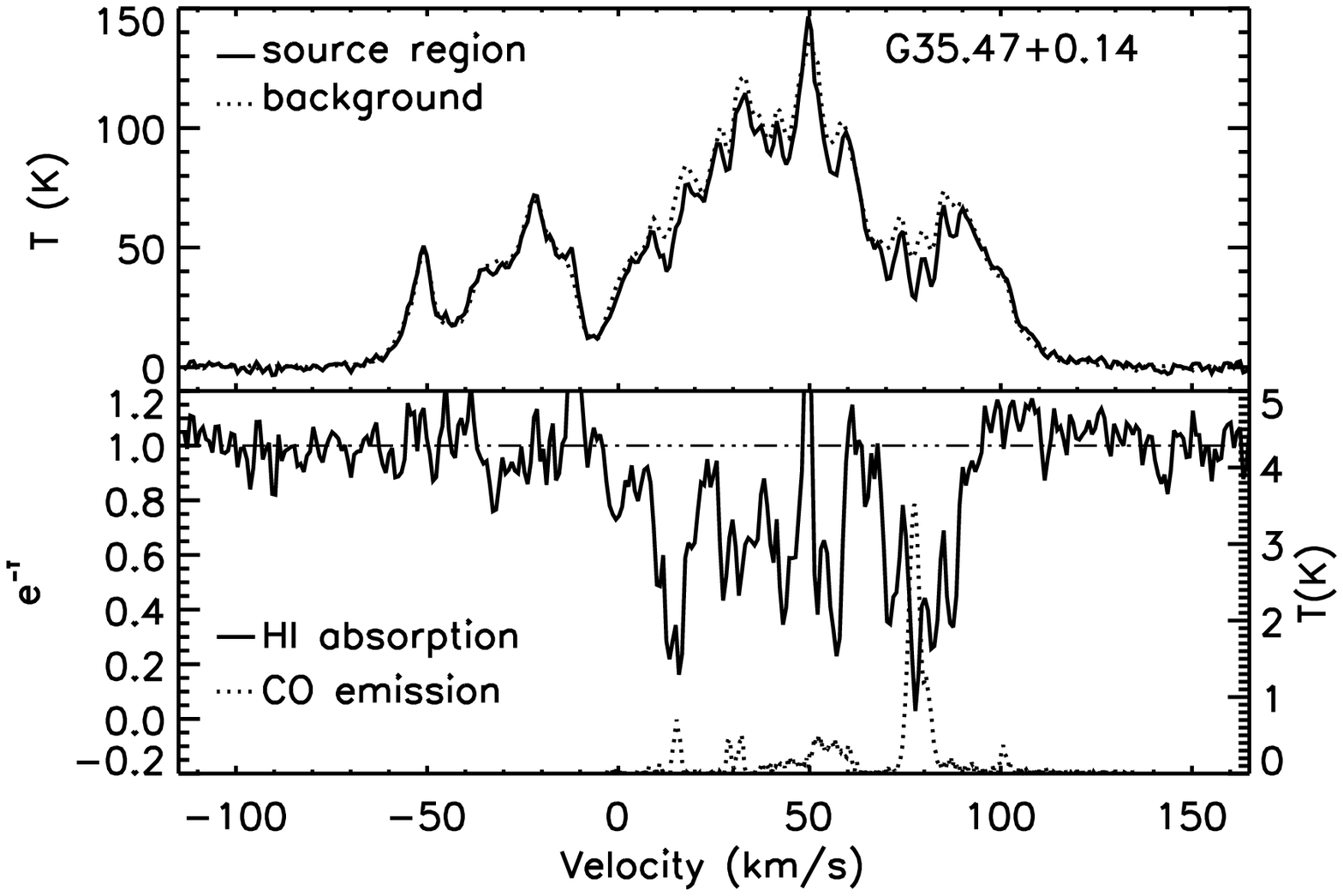}\includegraphics[width=0.5\textwidth, angle=0]{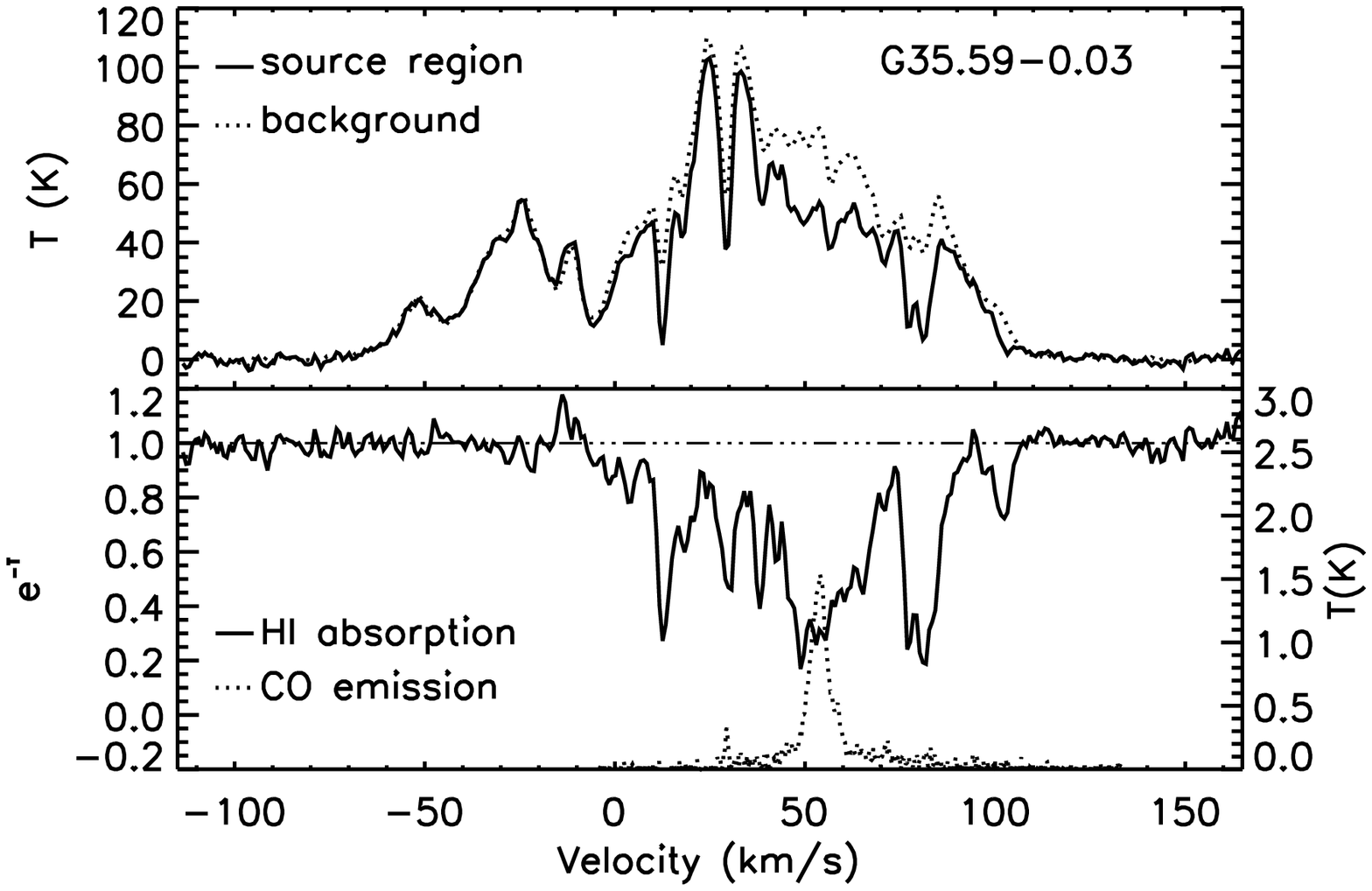}}\caption{The {\HI}, ${}^{13}\textrm{CO}$ emission and {\HI} absorption spectra of 4 background {\HII} regions: G35.05-0.52 (upper-left), G35.14-0.76 (upper-right), G35.47+0.14 (lower-left), G35.59-0.03 (lower-right). }
\label{fig:backgroundabsorption}
\end{figure*}

\begin{figure*}[!htpb]
\centerline{\includegraphics[width=0.5\textwidth, angle=0]{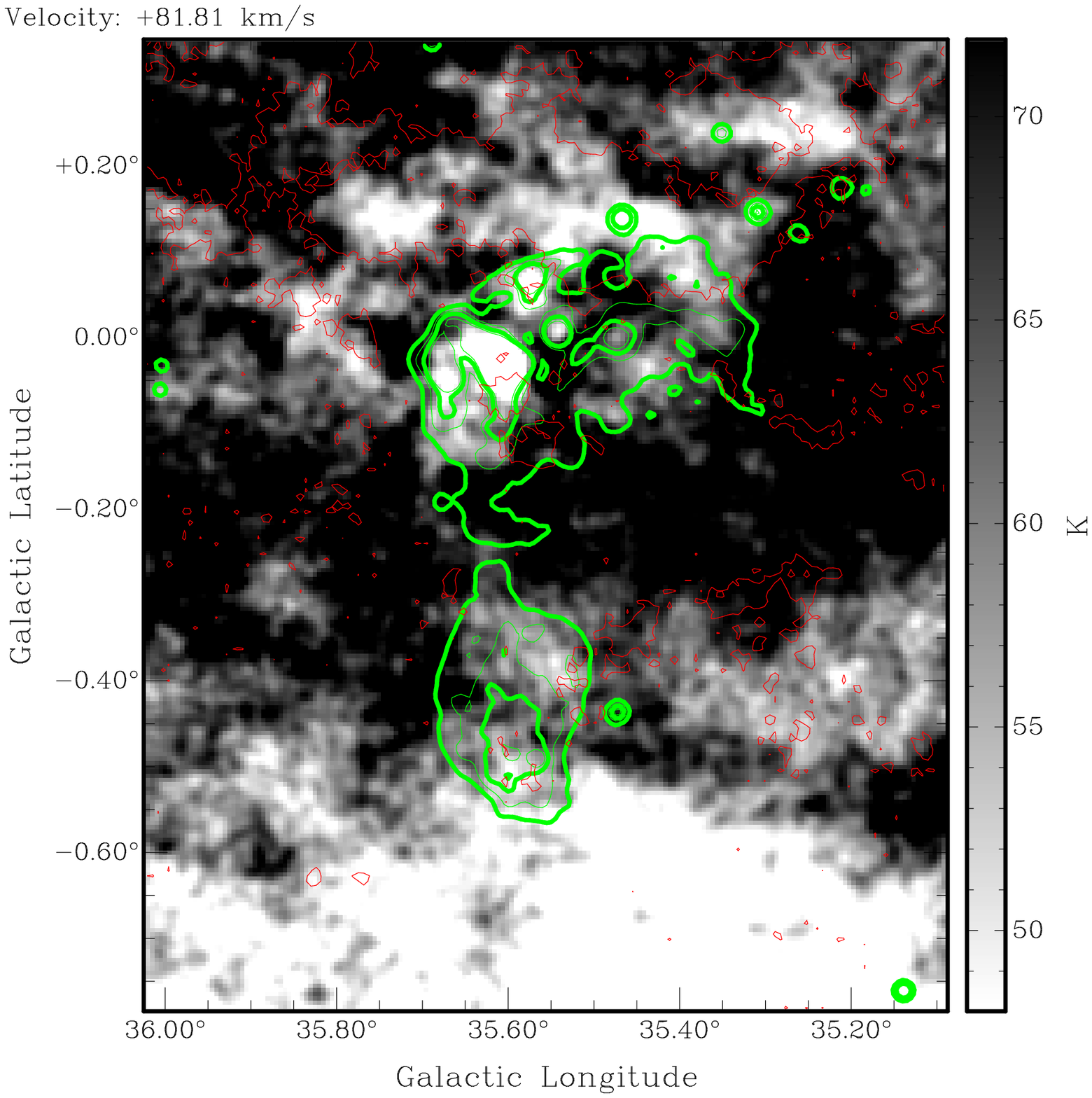}\includegraphics[width=0.5\textwidth, angle=0]{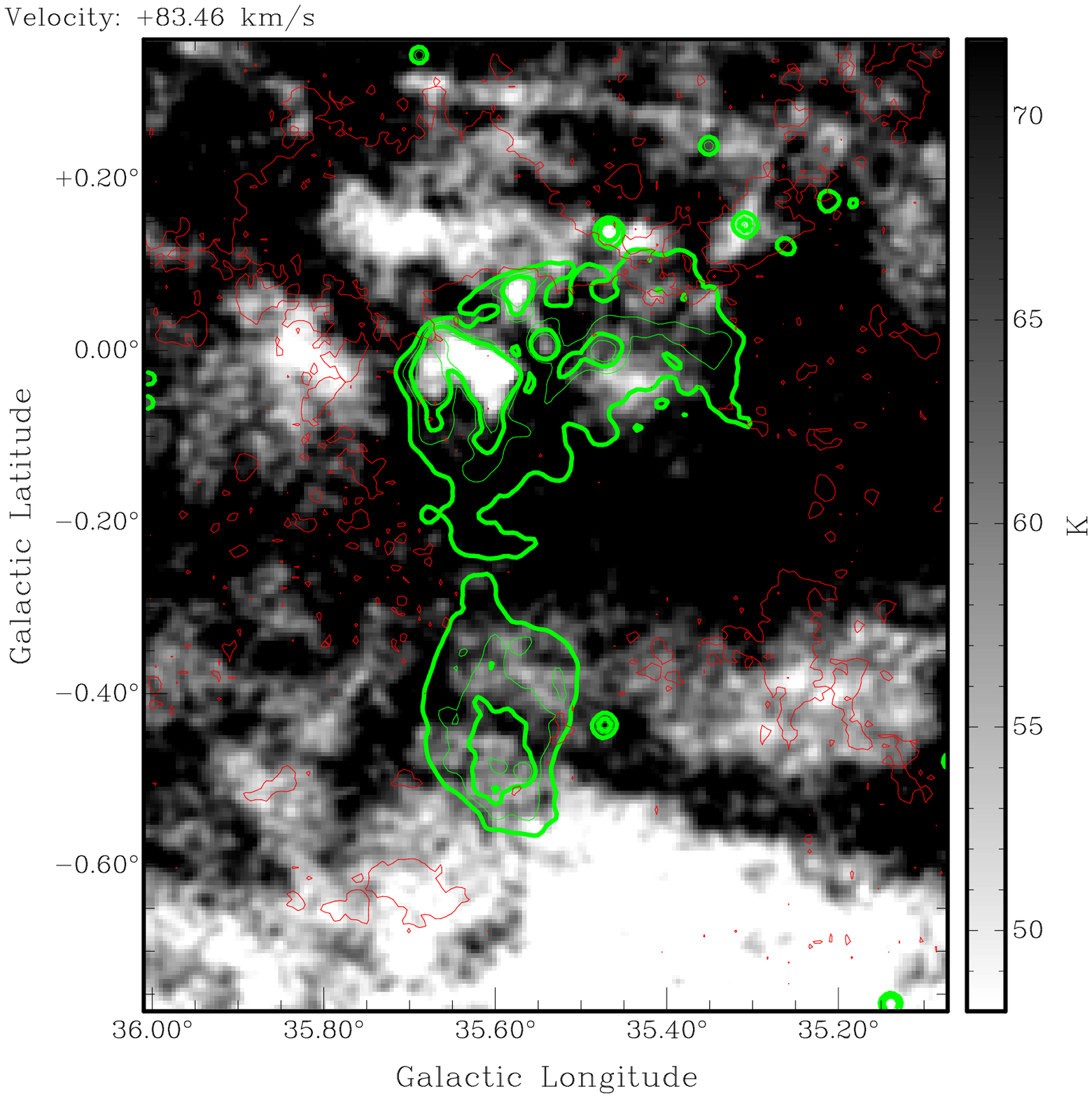}}
\centerline{\includegraphics[width=0.5\textwidth, angle=0]{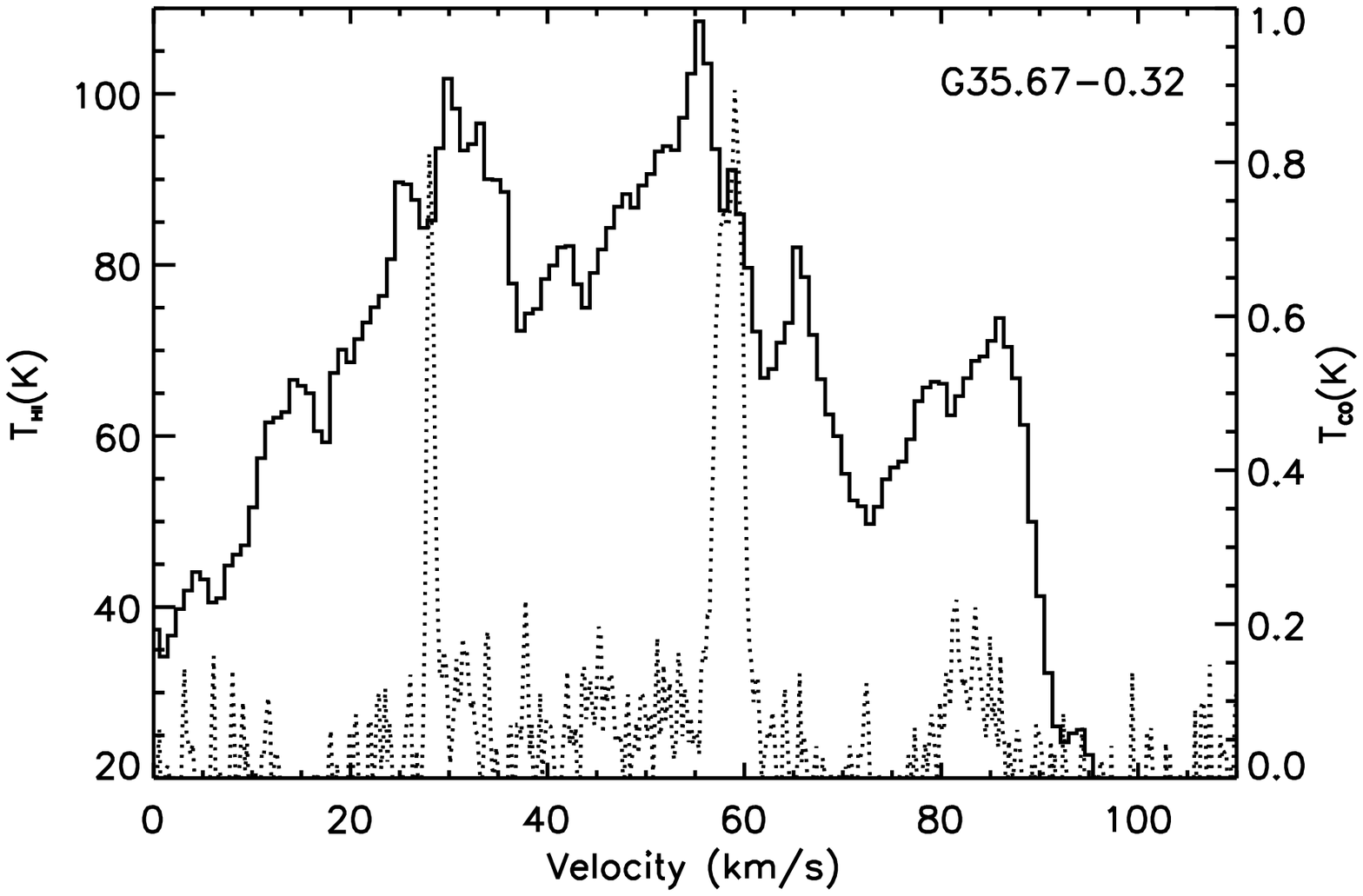}\includegraphics[width=0.5\textwidth, angle=0]{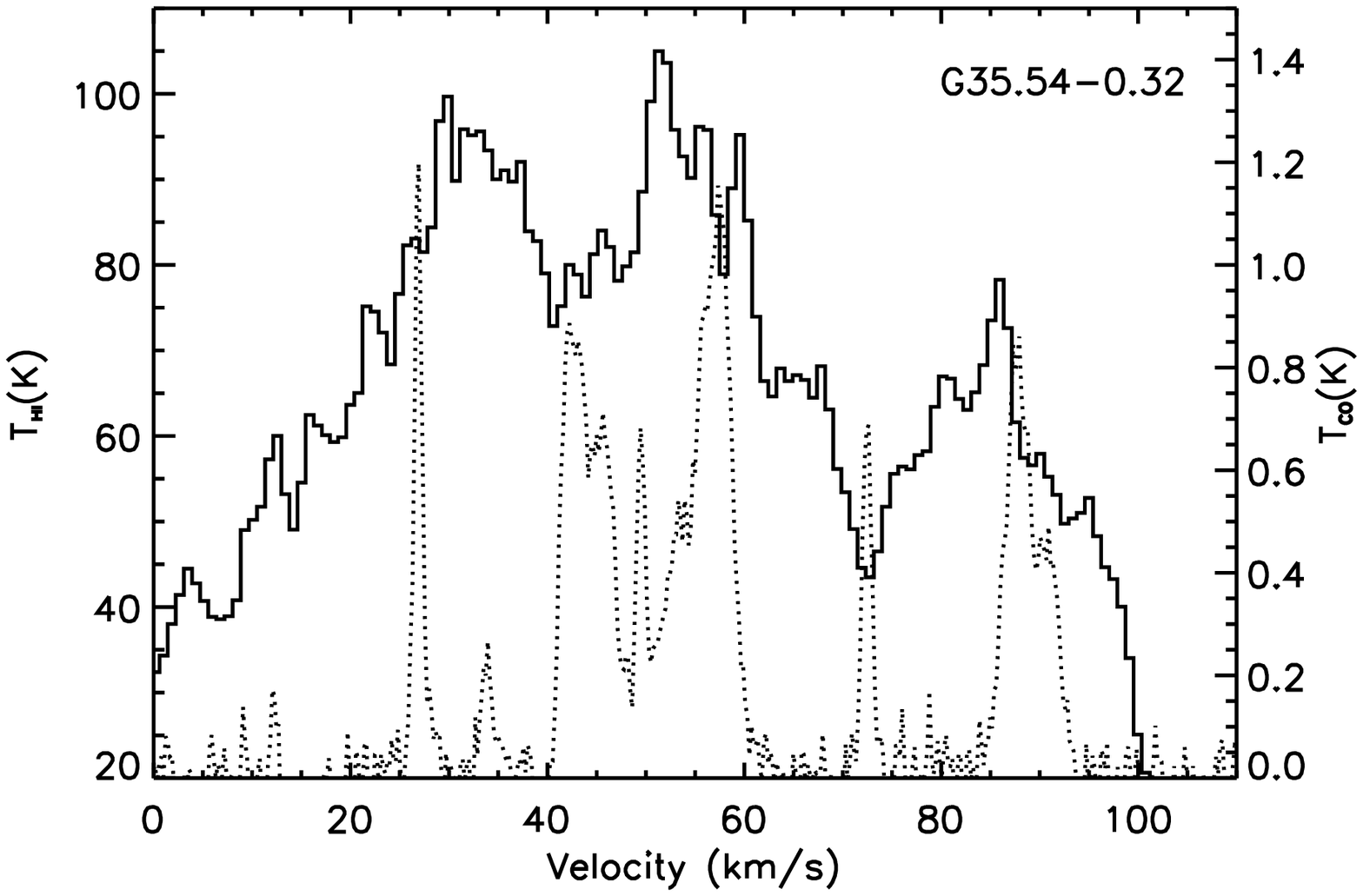}}
\centerline{\includegraphics[width=0.5\textwidth, angle=0]{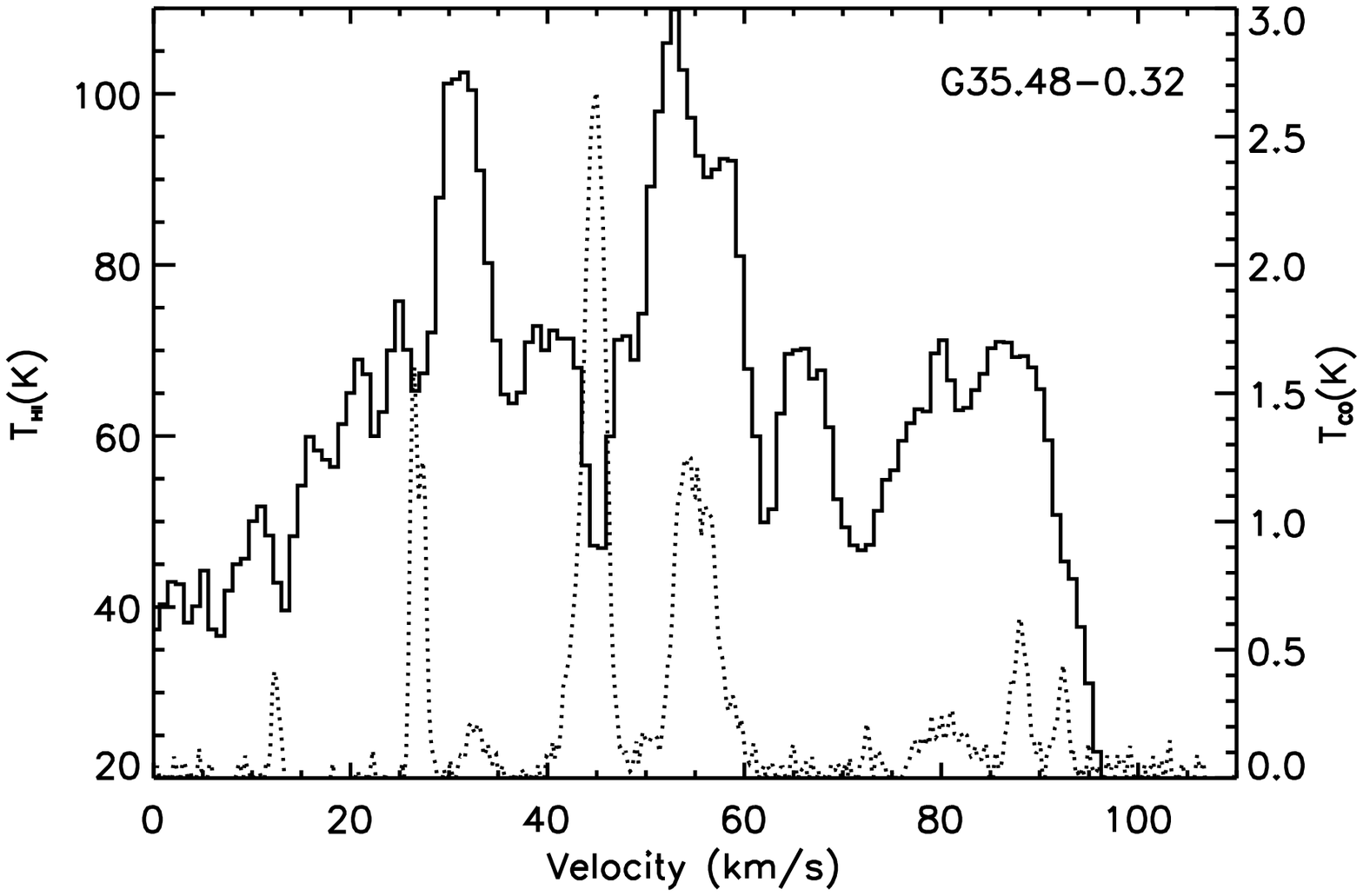}\includegraphics[width=0.5\textwidth, angle=0]{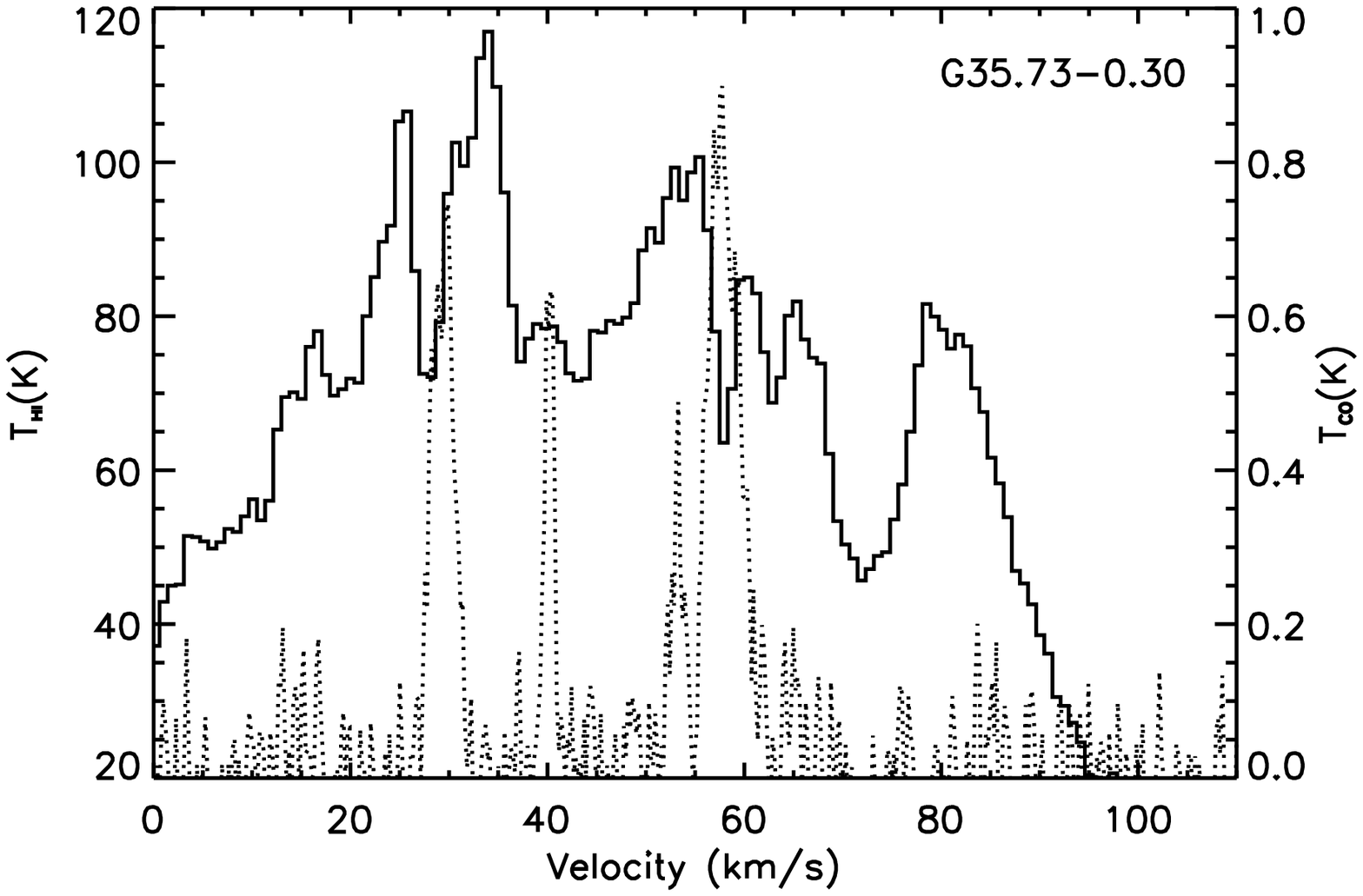}}\caption{Upper:{\HI} channel maps at 81.81 km s${}^{-1}$ (left) and 83.46 km s${}^{-1}$ (right) with overlay of 1420 MHz continuum contours from Figure \ref{fig:complex} (green) and ${}^{13}$CO J=1-0 emission contours with level of 0.3 K per channel (red). Middle and lower: The HI (histogram) and  ${}^{13}$CO J=1-0 emission (dot-line) at four regions around the complex.}

\label{fig:channelmap}
\end{figure*}

The absorption spectrum of PN G035.5-00.4 averaged over 2 adjacent velocity channels is shown in Figure \ref{fig:sourceabsorption}. The spectrum reveals that the highest absorption velocity is $\sim$68 km s${}^{-1}$. However, this feature disappears in the absorption spectrum averaged over 4 adjacent velocity channels (we do not show the spectrum in the article). It also cannot be found when we use the average of regions from 2 to 7 (see the right panel of Figure \ref{fig:complex}) as the background region to construct the absorption spectrum. Therefore, the 68 km s${}^{-1}$ feature is likely caused by small scale fluctuations of {\HI} clouds. It is more reasonable to take 58 km s${}^{-1}$ as the highest absorption velocity. The ${}^{13}\textrm{CO}$ emission peak shown at the same velocity confirms the reality of the absorption. This suggests a lower limit distance of 3.8$\pm$0.4 kpc to PN G035.5-00.4. The lack of {\HI} absorption at 82 km s${}^{-1}$ where there is a ${}^{13}\textrm{CO}$ emission peak suggests an upper limit distance of 5.4$\pm$0.7 kpc to PN G035.5-00.4.
\newline

\subsubsection{SNR G35.6-0.4 and {\HII} region G35.6-0.5}

\noindent The bottom panel of Figure \ref{fig:sourceabsorption} displays the absorption spectrum of SNR G35.6-0.4 and the {\HII} region G35.6-0.5. Two absorption features at velocity of $\sim -10$ km s${}^{-1}$ and $-46$ km s${}^{-1}$ correspond to distances of $\sim$14.7 kpc and $\sim$19.2 kpc, respectively. No absorption features are seen from 70 km s${}^{-1}$ to 100 km s${}^{-1}$ where HI self-absorption features are found around the complex. If the SNR lies behind those cold clouds, HI absorption should be seen just like the cases of other background sources, such as G35.59-0.03, G35.47+0.14, and G35.05-0.52. However, no absorption features are found in either the $\sim$82 km s${}^{-1}$ channels map or the HI absorption spectrum (see Figure 4). Therefore, large scale {\HI} variation across the ON-OFF region likely make the $-10$ km s${}^{-1}$ and $-46$ km s${}^{-1}$ absorption features.

\citet*{par10} found $\sim$55 km s${}^{-1}$ ${}^{13}$CO emission from the MC on the southern border of SNR G35.6-0.4. The emission line is asymmetric and has a slight broadening (see their Figure 2). They suggested that these features are the result of the interaction between the SNR and the MC. Then, the MC and the SNR G35.6-0.4 should be at the same distance, i.e. 3.6$\pm$0.4 kpc at the near side or 10.2$\pm$0.4 kpc at the far side. The absorption spectrum of the bottom panel of Figure \ref{fig:sourceabsorption} shows absorption features up to $\sim$61 km s${}^{-1}$, which suggests that the near side distance of 3.6$\pm$0.4 kpc is plausible.

\citet*{loc89} observed a narrow RRL (the width is 28.9$\pm$3.6 km s${}^{-1}$) at 56.0$\pm$2.6 km s${}^{-1}$ towards G35.6-0.5, while \citet*{phi93} found a broad RRL (its width is 39$\pm$5 km s${}^{-1}$) at 54$\pm$1 km s${}^{-1}$ in the same direction. These authors believed that the RRL originated from the {\HII} region G35.6-0.5. \citet{par11} found clear polycyclic aromatic hydrocarbons emission at 8 $\mu$m, which suggests G35.6-0.5 is likely an extended {\HII} region. The RRL peak antenna temperature (${T_l}$) of of \citet{phi93} is $91\pm5$ mK, which implies a relative signal to noise ratio of 18.2. In the article of \citet{loc89}, $T_l$ is equal to $24 \pm 2.3$ mK and the calculated relative signal to noise ratio is 10.4.

\citet*{phi93} found that the full-width at half maximum (FWHM) of the RRL is nearly 2 times greater than the expected value from thermal broadening for a $10^{4}$ K {\HII} region. This implies enhanced turbulence caused by a non-thermal source. Therefore, SNR G35.6-0.4 may be adjacent to the {\HII} region G35.6-0.5. \citet*{gre09} suggested that the observed RRL may originate not from the {\HII} region G35.6-0.5 but from the PN G035.5-00.4, since PN G035.5-00.4 is very close to G35.6-0.5. A spectroscopic observation with high angular and spectral resolution is needed to solve this question. If the RRL partly comes from the {\HII} region G35.6-0.5, this would hint that the source is located at a distance of 3.6$\pm$0.4 kpc. Table \ref{tbl:summary} summarizes the velocities of HI absorption, CO emission features and distances of the studied objects.

\begin{figure*}[!htpb]
\centerline{\includegraphics[width=0.35\linewidth, angle=270]{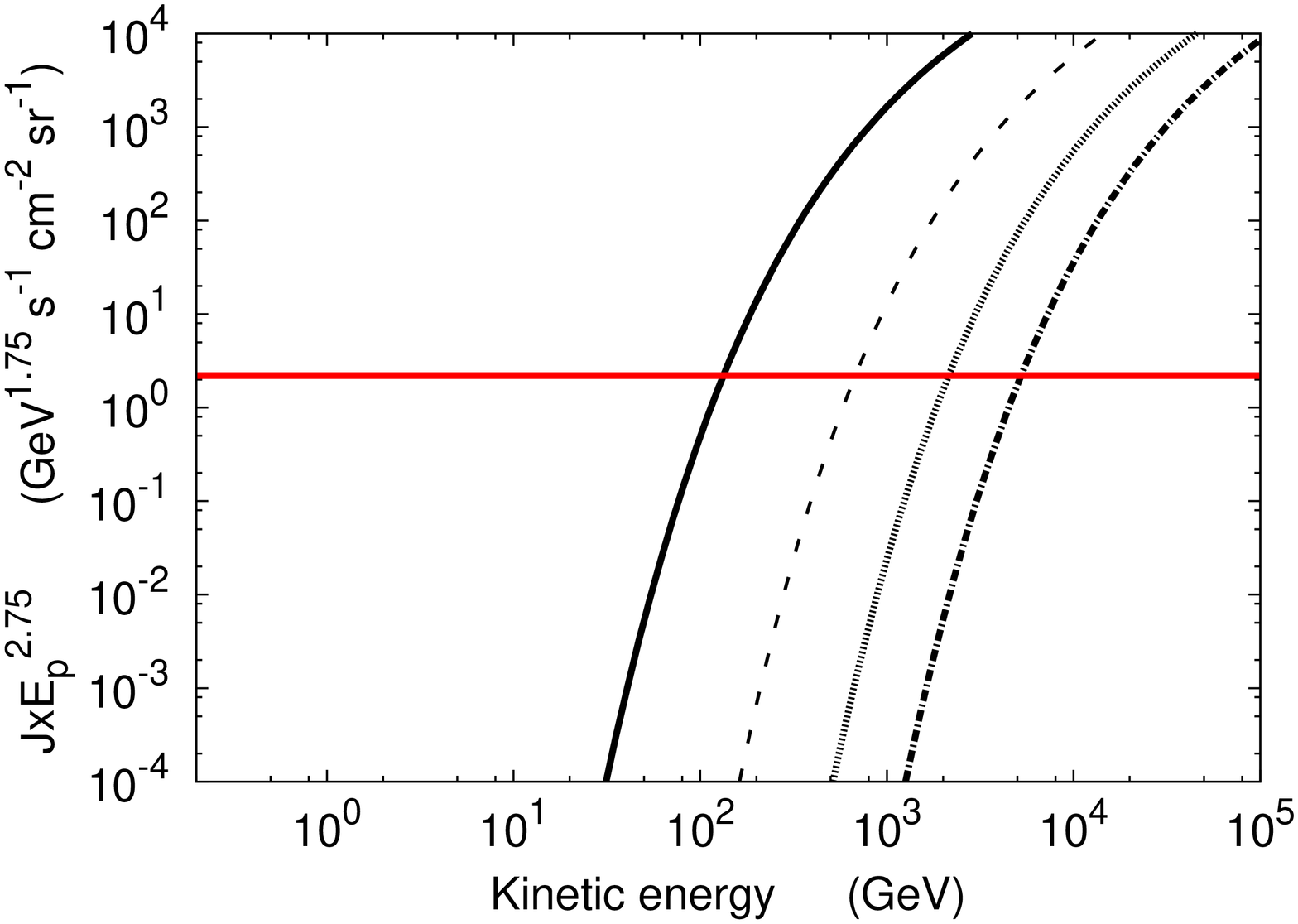} \includegraphics[width=0.35\linewidth, angle=270]{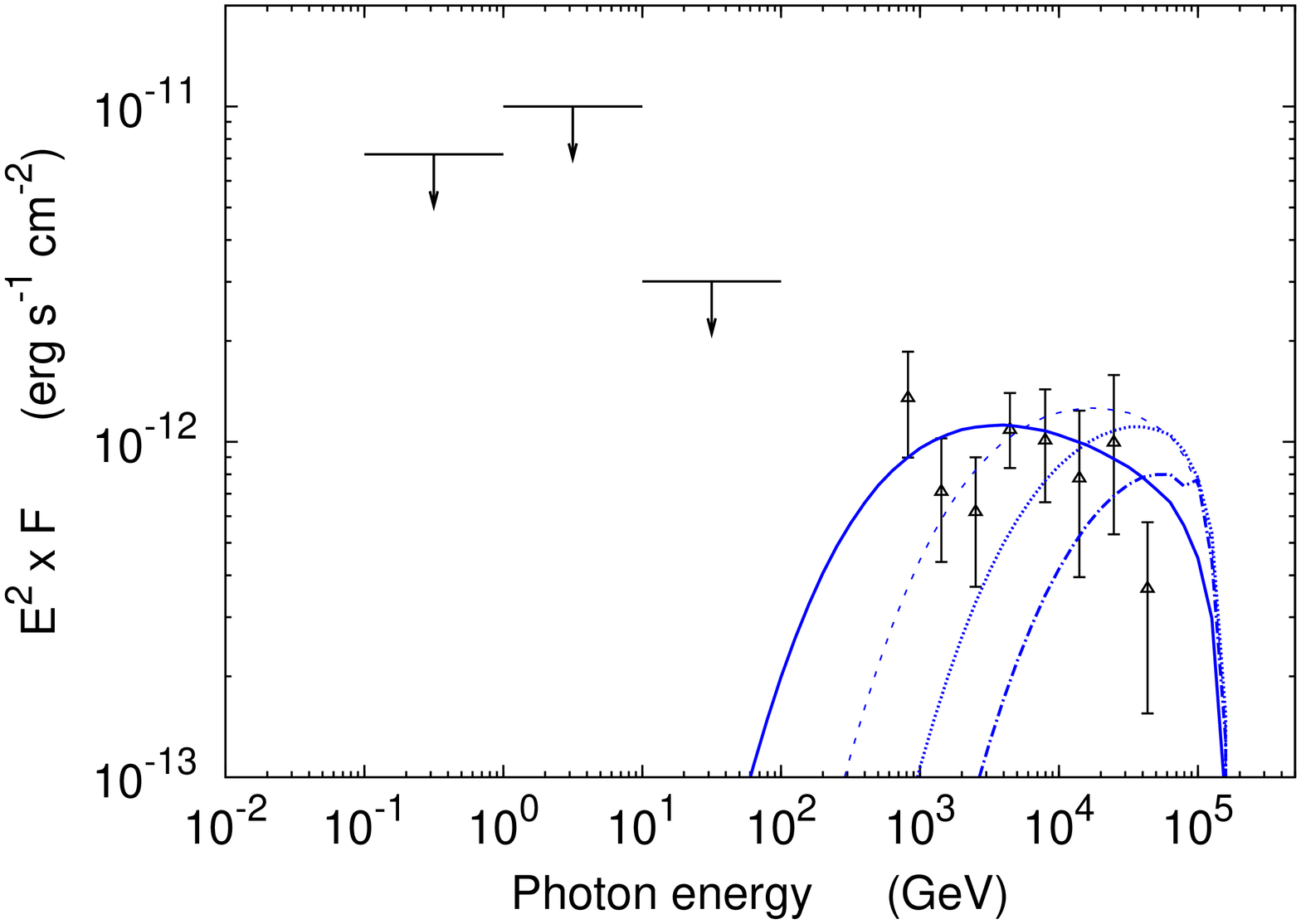}}
\centerline{\includegraphics[width=0.35\linewidth, angle=270]{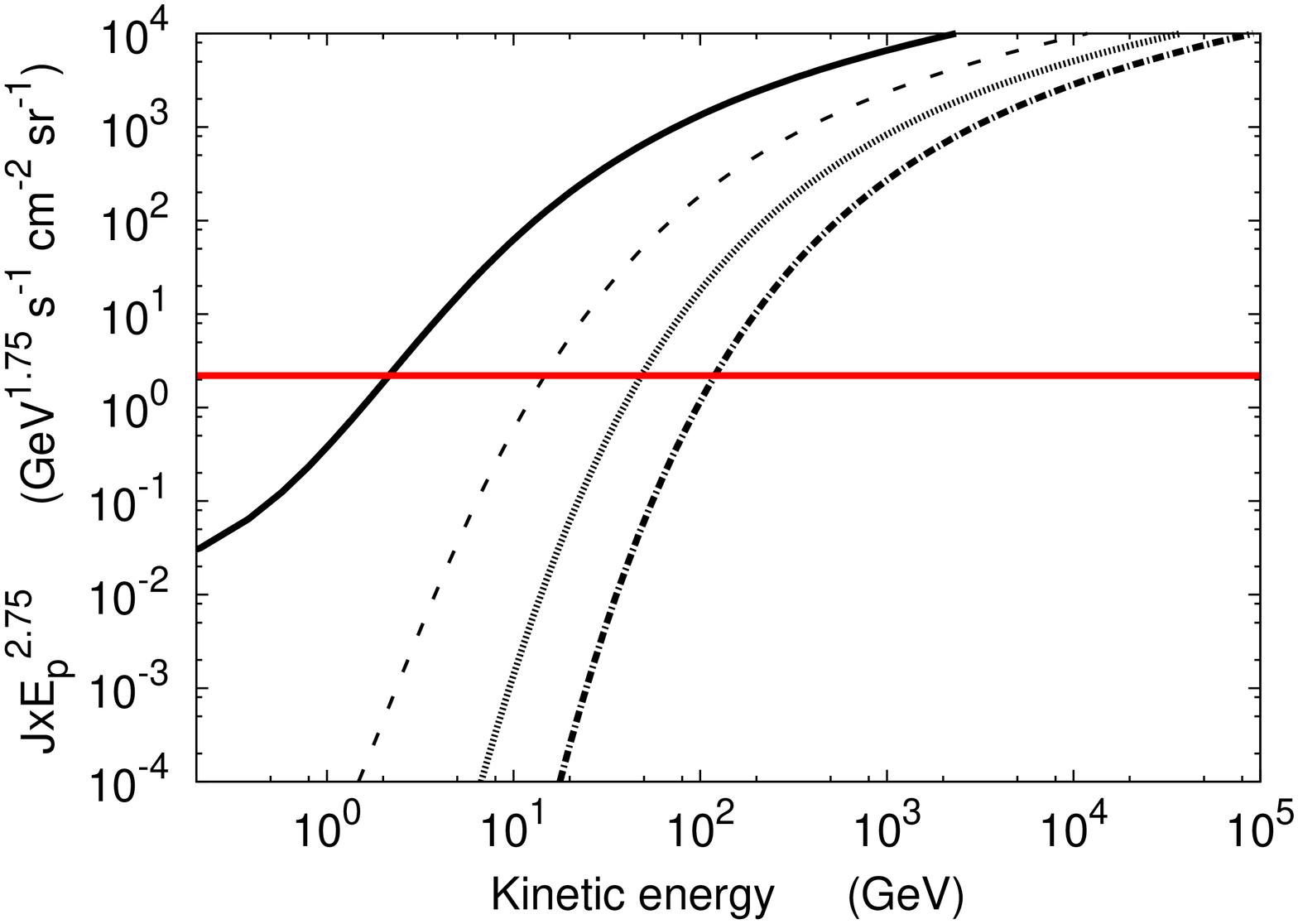} \includegraphics[width=0.35\linewidth, angle=270]{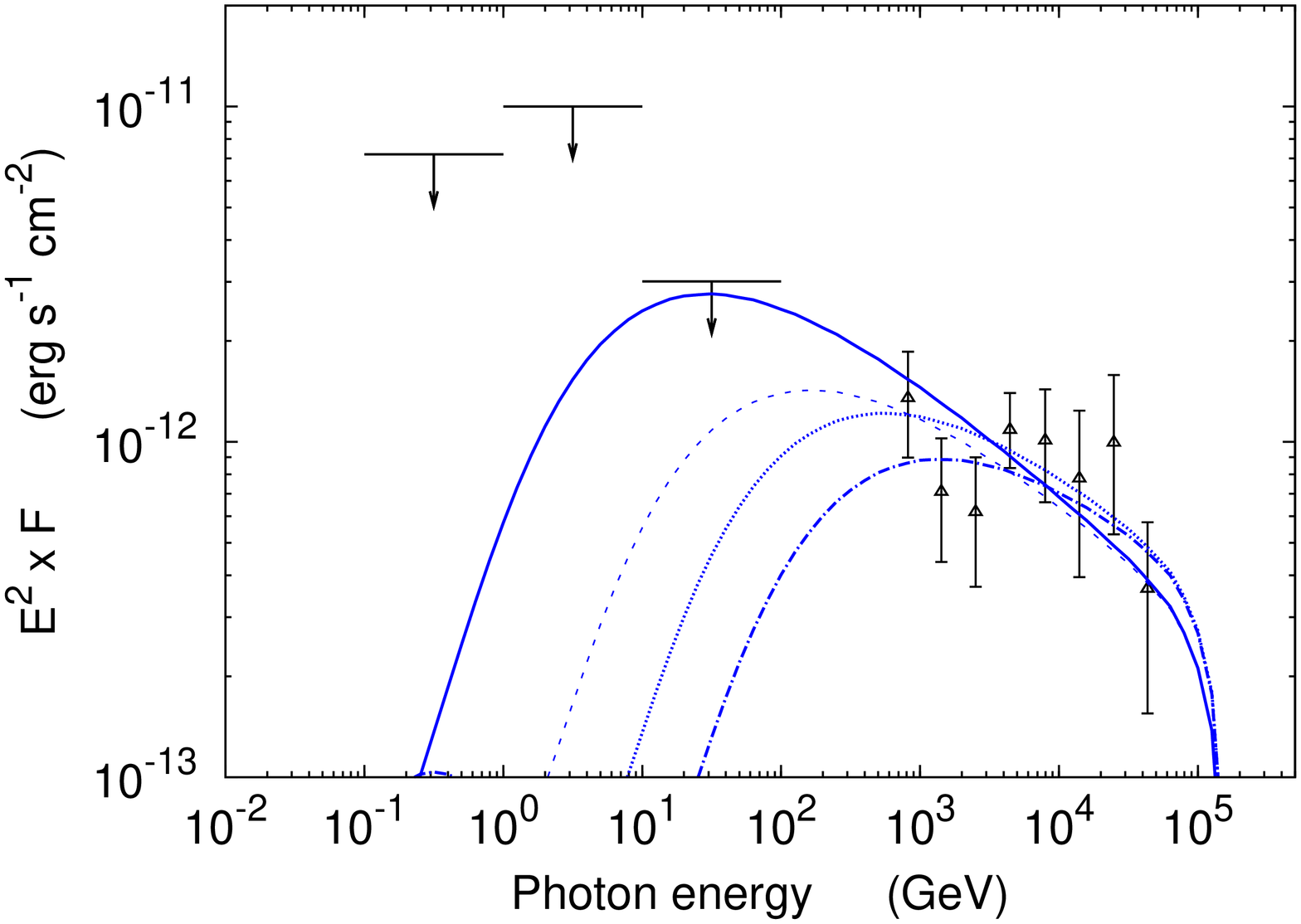}}
\centerline{\includegraphics[width=0.35\linewidth, angle=270]{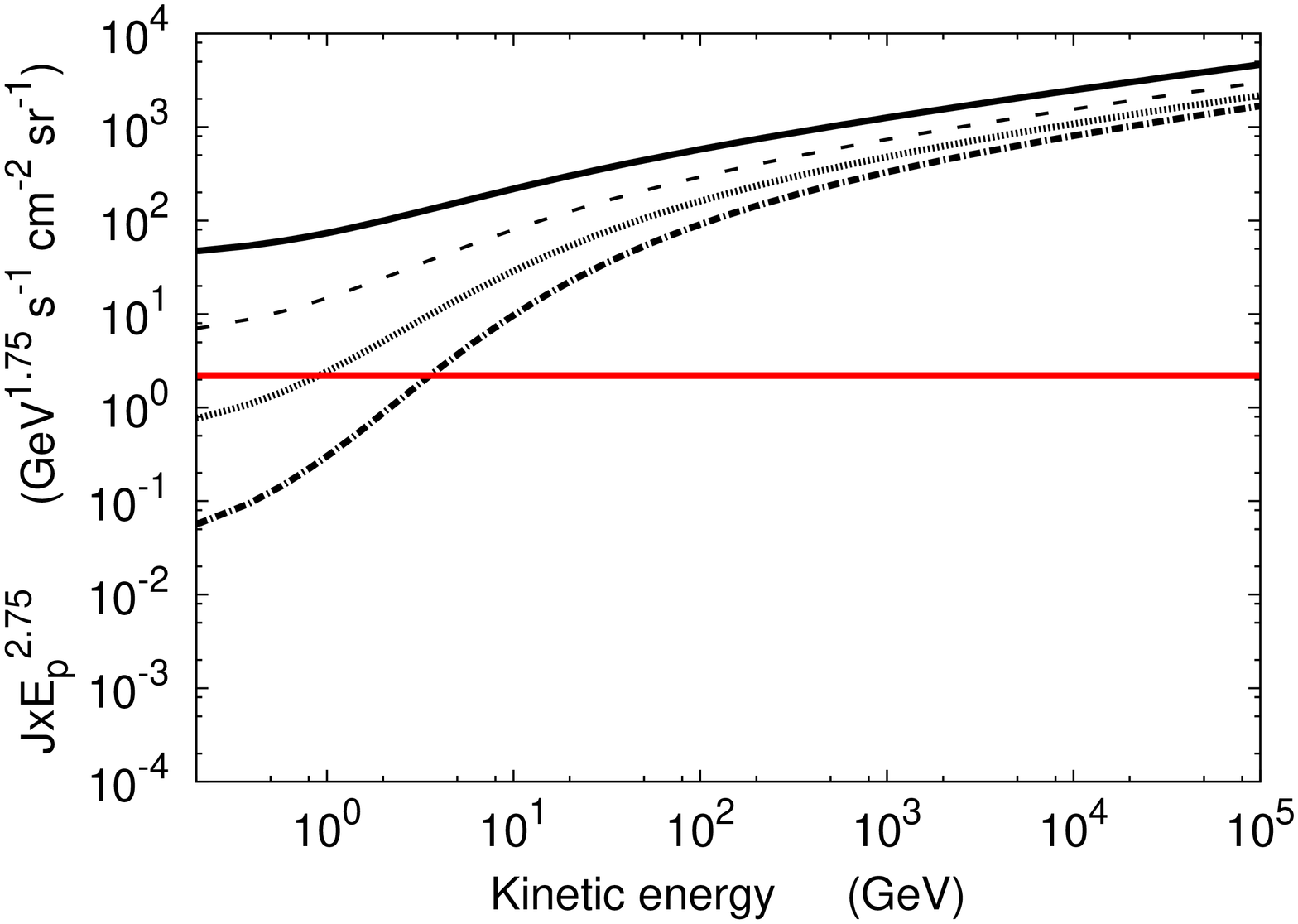} \includegraphics[width=0.35\linewidth, angle=270]{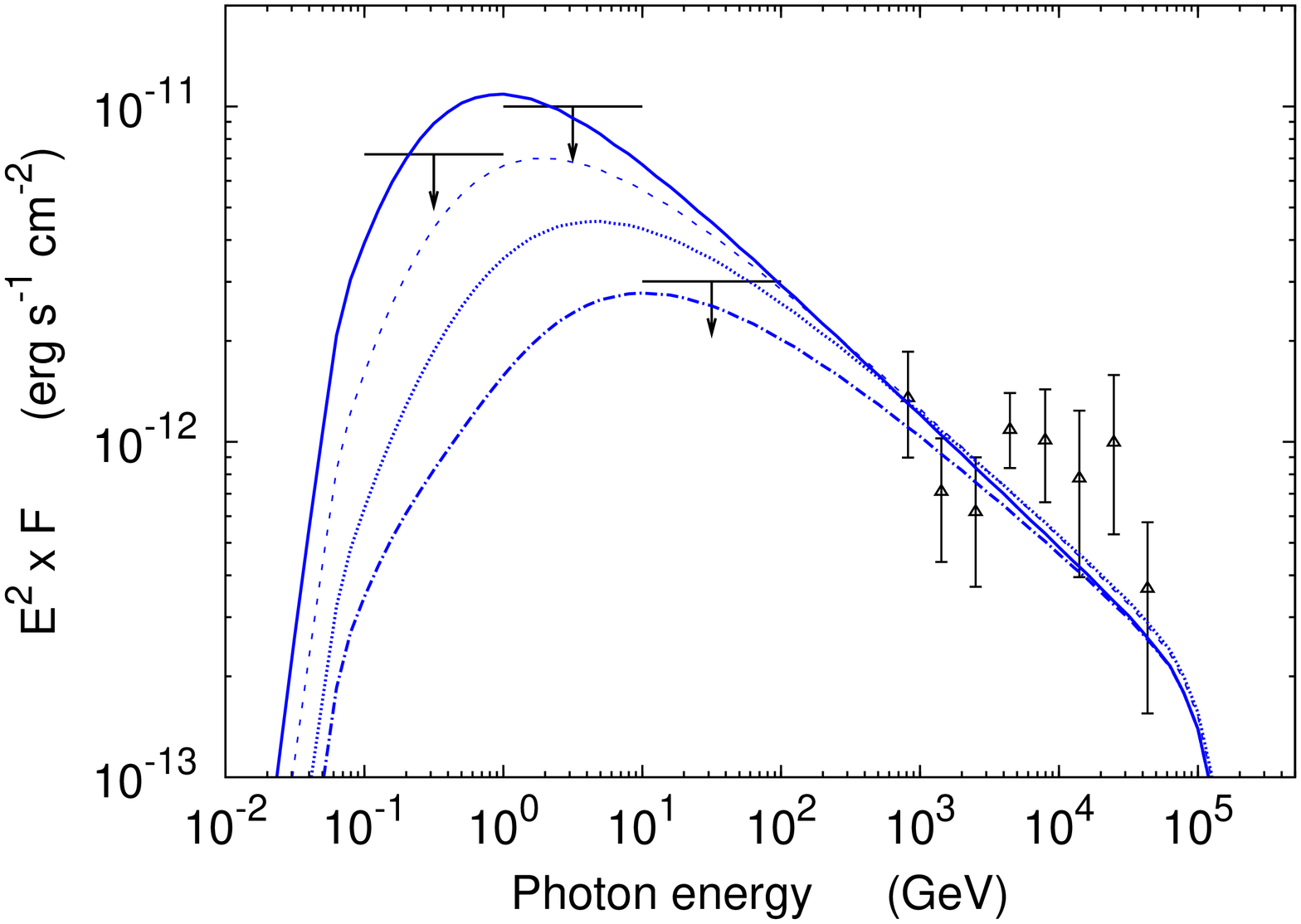}}
\caption{Point-like injection models for the gamma-ray spectrum of SNR G35.6-0.4. The parameters used in this figure correspond to those given in Table \ref{tbl:param_models} and the distance to the SNR/MC complex as proposed in this work. The solid, dashed, dotted and dot-dashed curves represent models with increasing values of separation radius, as given in Table \ref{tbl:param_models}. The left panels show the differential proton flux compared to the cosmic-ray sea (solid red line). The right panels show the fitting models of the SED, compared to H.E.S.S. observations \citep{aha08} and {\it Fermi}-LAT upper limits \citep{tor11}. From left to right, the assumed normalization of the diffusion coefficient is $D_\mathrm{10}=10^{26, 27, 28}$ cm$^{2}$ s$^{-1}$.}
\label{fig:fittingmodels}
\end{figure*}

\section{Discussion}

\subsection{{\HII} region G35.47+0.14 and G35.14-0.76}

\noindent \citet{kol02} identified G35.47+0.14 as an extragalactic continuum source whose absorption spectrum shows {\HI} absorption at negative velocities less than -10 km s${}^{-1}$. However, G35.47+0.14's spectrum in the lower-left panel of Figure 4 shows absorptions neither at the tangent point velocity nor at the negative velocity. Considering the detected RRL at 80.9$\pm$0.5 km s${}^{-1}$, G35.47+0.14 is likely at a distance of 5.3$\pm$0.7 kpc. In addition, the 6 cm continuum observation \citep{urq09}, mid-infrared observation \citep{giv07} and {\HI} self-absorption \citep*{and09} also support G35.47+0.14 is a Galactic object.

\citet{fro11} studied G35.14-0.76 and found the central star has a spectral type of O9 or B0 and its mass is about 20 $\Msol$. They suggested that G35.14-0.76 is a misclassified PN and should be re-classified as an {\HII} region. The 2MASS K-band extinction to G35.14-0.76 is 0.78 mag which indicates a distance of 2.5 kpc to G35.14-0.76. This distance consists with the kinematic distance of 2.4$\pm$0.5 kpc. \citet{han11} detected a H${}_{2}$CO feature at 33.7$\pm$0.2 km s${}^{-1}$ nearly at the same velocity as the highest {\HI} absorption velocity, $\sim$35 km s${}^{-1}$. Therefore 2.4$\pm$0.5 kpc is a feasible distance estimate to {\HII} region G35.14-0.76.

\subsection{IRAS 18551+0159 and PN G035.5-00.4}

\noindent IRAS 18551+0159 has been observed by Midcourse Space Experiment (MSX) satellite \citep{ega03}. Its 8.28 $\mu$m flux density is 1.47 Jy. According to the mid-infrared distance scale \citep{ort11}, \[\log {D} = {-0.1736}\log {F_{8.28um}} - {0.3899}\log \theta  + {0.7960}\] the calculated statistical distance of IRAS 18551+0159 is 5.0 kpc. Considering the large error in statistical method, this distance consists with the kinematic distance, 4.3$\pm$0.5 kpc. \citet{ede88} detected OH emission (1612 MHz) from IRAS 18551+0159, which reveals that its age is less than $\sim$1000 yr \citep{gom07}. \citet{jim05} found IRAS 18551+0159 is extremely red and has no optical counterpart. This indicates IRAS 19551+0159 may has departed from the AGB recently. With the distance of 4.3$\pm$0.5 kpc, the size of IRAS 18551+0159 is about 0.06 pc. The small size supports that IRAS 18551+0159 left the AGB phase recently.

For PN G035.5-00.4, the distance estimate (from 3.8$\pm$0.4 kpc to 5.4$\pm$0.7 kpc) is similar to the statistical distance, 4.6 kpc \citep{ort11}. The new distance restriction suggests a normal linear-size of about 0.2$\sim$0.3 pc for PN G035.5-00.4. \citet{coh07} studied PN G035.5-00.4 based on the {\it Spitzer}/IRAC data and found that its morphology changes from asymmetric enhanced brightness of the southern limb in the three shortest bands (3.6, 4.5 and 5.8 $\mu$m) to a large circular appearance at 8 $\mu$m. This character can be explained by a substantial photodissociation region that envelops the entire ionized zone. Although the kinematic distance measurements does not rule out the possibility of an association with SNR G35.6-0.4 or {\HII} region G35.6-0.5, the perfect circle pattern implies no significant interaction between PN G035.5-00.4 and SNR G35.6-0.4 or {\HII} region G35.6-0.5.

\citet{and12} derived a robust color criterion to distinguish between {\HII} regions and PNe. They found nearly 98$\%$ of {\HII} regions in their sample have the 12 $\mu$m and 8$\mu$m flux density ratio less than 0.3. According to the data from MSX \citep{ega03}, this ratio is 2.4 and 3.4 for IRAS 18551+0159 and PN G035.5-00.4, respectively. The values support both IRAS 18551+0159 and PN G035.5-00.4 are likely real PNe.

\subsection{SNR G35.6-0.4}

\noindent With a distance of 3.6$\pm$0.4 kpc, the average size and age of SNR G35.6-0.4 would be revised to about 15 pc in diameter and 2300 yr. The new age would imply that the SNR G35.6-0.4 is in its early evolution stage. \citet*{phi93} suggested that SNR G35.6-0.4 is associated with PSR J1857+0212 because both objects have similar distances. However, the new kinematic distance to SNR G36.5-0.4 (3.6$\pm$0.4 kpc) is much less than the DM distance of 7.98 kpc to PSR J1857+0212. Furthermore, \citet{cli88} measured the HI absorption spectrum of PSR J1857+0212. The spectrum displays absorption features up to the tangent point velocity (see their Figure 4 for detail), which indicates a lower limit distance of 6.9$\pm$1.3 kpc, to PSR J1857+0212. This suggests PSR J1857+0212 could at least be 3 kpc behind SNR G35.6-0.4. \citet{mor02} discovered another pulsar PSR J1857+0210 near the center of SNR G35.6-0.4. However, PSR G1857+0212 has a DM distance of 15.4 kpc. An association between the pulsar and SNR G35.6-0.4 seems impossible. Moreover, both pulsars have large characteristic ages, which also disfavor the association (the characteristic age of PSR J1857+0212 and PSR J1857+0210 is about 160000 and 712000 years, respectively). Although the characteristic age is not the real age, which is usually unknown, it could be a good estimator of the age when the braking index is $\sim$3 and the initial pulsar spin-down period is much shorter than the current one.

Both pulsars have a relatively low spin-down power, e.g. $2.2 \times 10^{34}$ erg s${}^{-1}$ for PSR J1857+0212. Even at the distance we estimated for G35.6-0.4 (i.e., assuming that its DM distance is underestimated by more than a factor of 2), $\dot E / 4\pi D^2$ would be $1.3 \times10^{32}$ erg s${}^{-1}$ kpc${}^{-2}$. At typical efficiencies for radiation in the gamma-ray band, pulsar wind nebula driven by these pulsars would be undetectable by the current generation of instruments \citep[see e.g.,][]{mar12}. We can safely consider that none of the pulsars significantly contributes to the TeV source detected by H.E.S.S.

Does the weak gamma ray source HESS J1858+020 originate from the interaction between SNR G35.6-0.4 and the $\sim$55 km s${}^{-1}$ MC? \citet*{par10} and \citet{tor11} gave somewhat different opinions on the question. \citet*{par10} believed SNR G35.6-0.4 is interacting with the cloud at $\sim$ 55 km s${}^{-1}$ crossing the SNR shell. They found the cloud shows some possible kinematical evidence of shocked gas in ${}^{13}\textrm{CO J=1-0}$ emission spectrum: asymmetric and a slight spectral line broadening (see their Figure \ref{fig:sourceabsorption}). Assuming the clouds' distance of 10.5 kpc, the same as the SNR, they estimated its mass and density to be $\sim$5$\times 10^{3}$ $\Msol$ and $\sim$500 cm$^{-3}$, respectively, which seem enough to explain the observed gamma ray flux. \citet{par11} excluded the possibility that HESS J1858+020 originates from molecular outflows of a young stellar object. Therefore, SNR G35.6-0.4/$\sim$55 km s${}^{-1}$ MC becomes the most probable counterpart of HESS J1858+020.

\citet{tor11} have considered a possible association between G35.6-0.4 and HESS J1858+020 doubtful because of the lacking of a GeV counterpart. Indeed, they have analyzed 2 years of Fermi-LAT data for the region of interest, and considered whether it was possible that the closest LAT source, 1FGL J1857.1+0212c, could be spatially related to HESS J1858+020. Concluding it is not, at least with dataset at hand, upper limits were imposed. Thus, the cosmic-ray interaction between protons accelerated in the SNR and material in the cloud should be such to allow producing the TeV emission of HESS J1858+020 without producing a GeV counterpart. This may in principle be possible for particular combinations of diffusion coefficients and MC/SNR distances (low mass in the cloud and slow diffusion timescales). In fact, this very scenario has been claimed for other sources, like SNR W28 \citep*[e.g.][]{li10}.

\begin{table}[t]
\centering
\caption{Parameters for point-like continuous injection as shown in Figure \ref{fig:fittingmodels}.}
\begin{tabular}[t]{ccc}
\hline
$D_\mathrm{10}$  & $R_\mathrm{SNR/MC}$    &    Mass   \\
cm$^{2}$ s$^{-1}$ & (pc)   &   $(\Msol)$   \\
\hline
$10^{26}$&10&150\\
$10^{26}$&15&500\\
$10^{26}$&20&1000\\
$10^{26}$&25&1500\\
\hline
$10^{27}$&10&500\\
$10^{27}$&15&800\\
$10^{27}$&20&1500\\
$10^{27}$&25&2000\\
\hline
$10^{28}$&10&3000\\
$10^{28}$&15&5000\\
$10^{28}$&20&7000\\
$10^{28}$&25&8000\\
\hline
\end{tabular}
\label{tbl:param_models}
\end{table}

\citet{tor11}'s calculation was based on a distance to the SNR/MC complex of 10.5 kpc, and a molecular mass of several thousand solar masses. Accepting the new distance estimate for the remnant, of 3.6$\pm$0.4 kpc, suggests much less mass is available for cosmic-ray interactions, of the order of 600 $\Msol$. Assuming these parameters, we have reconsidered the model put forward by \citet{tor11} and the results are as follows. For details regarding the numerical model itself we refer the reader to the works by \citet{rod08}, and \citet{tor08, tor10}

The physical size of the SNR shell and the projected separation to the MC are also resized due to the revised distance to the SNR/MC complex, as seen above. This means that the accelerated particles diffuse to much shorter distances than what was considered before. The maximum of the particle flux, for a given energy $\Ep$ at a given time $t$, is reached at the distance $R=\sqrt{6 \ t \, D(\Ep)}$ \citep*[see, e.g.,][]{aha96}. This is equal to a projected distance of $\sim 10$ pc for energies of $\Ep\sim 5$ TeV and a diffusion coefficient of $D(E_\mathrm{p})=D_\mathrm{10} (E_\mathrm{p}\, / \, 10\,\textrm{GeV})^{\delta}$, with $D_\mathrm{10}=10^{26}$ cm$^{2}$ s$^{-1}$and $\delta$=0.5. The bulk of the particles with lower energies than 5 TeV, then, still have not reached the target mass. This is exact for an impulsive source, when the age of the source is much longer than the time over which the particles are injected.

Due to smaller age, the source is probably better approximated as a continuous injection point, as the short age is comparable to the time of release of the particles. The injection luminosity is such that the total energy input in the lifetime of the SNR is $10^{50}$ erg, hence equal to $1.38 \times 10^{39}$ erg s$^{-1}$. We choose the slope of the injection spectrum to match the one in \citet{tor11}, with $p=2.0$. The only parameters that are left free for a fit are then the actual separation of the accelerator (that would probably not exceed the projected separation by more than a factor of few) from the target and the actual mass of the target that interacts with the cosmic-ray population. Note that in \citet{tor11}, distances from the injection point in the shell were quoted. Here we simply refer to distances from the center of the SNR.

Figure \ref{fig:fittingmodels} shows the prediction from models having the parameters specified in Table \ref{tbl:param_models}. The cosmic-ray sea contribution is the dominant one at the lowest energies (in the {\it Fermi}-LAT regime). But, due to the small mass of the target cloud, the contribution to the flux given by the emission related to the cosmic-ray sea amounts to a very low flux, below the range of fluxes shown in the figures and is then invisible in there.

The results of Figure \ref{fig:fittingmodels} and Table \ref{tbl:param_models} show that the new distance and MC estimation does not rule out the possibility of a connection between G35.6-0.4 and HESS J1858+020. Actually, it is currently more favorable than in the case of a 10.5 kpc-distance, since the low values of diffusion coefficient required before are no longer needed to maintain the GeV luminosity below observational sensitivity. We note, however, that for a diffusion coefficient of the order of the galactic average one, i.e. $D_\mathrm{10}=10^{28}$ cm$^{2}$ s$^{-1}$, we need a large separation between the accelerator and the target, with a mass of the latter that largely exceeds the estimates given here by one order of magnitude. Therefore we still expect a suppressed diffusion coefficient in the environment in case of a physical association. We also note that the SNR diameter is of the order of 15 pc, so assuming 10 pc radius from the injection point is a bare minimum for a realistic model, and slightly higher separations would be preferred.

\begin{acknowledgements}
HZ, WWT and HQS acknowledge supports from NSFC (011241001, 211381001, Y211582001) and BaiRen programme of the CAS (034031001). This work is partly supported by China's Ministry of Science and Technology under State Key Development Program for Basic Research (2012CB821800, 2013CB837901). DFT and GP research has been done in the framework of the grants AYA2012-39303, as well as SGR2009- 811, and iLINK2011-0303. DFT was additionally supported by a Friedrich Wilhelm Bessel Award of the Alexander von Humboldt Foundation. This publication was partly supported by a grant from the John Templeton Foundation and National Astronomical Observatories of the CAS. The opinions expressed in this publication are those of the authors do not necessarily reflect the views of the John Templeton Foundation of NAOCAS. The funds from John Templeton Foundation were awarded in a grant to The University of Chicago which also managed the program in conjunction with NAOCAS. We thank Drs. L.Z. Wang, X.H. Cui, and Ms. D. Wu for meaningful discussions. We also thank the referee for suggestions which improved this paper.\\
\end{acknowledgements}

\bibliographystyle{apj}

\end{document}